\documentstyle[emulateapj,epsf]{article}

\received{}
\revised{}
\accepted{}
\cpright{}{}
 
\journalid{}{}
\articleid{}{}
\paperid{}
 
\cpright{}{}
\ccc{}
 
\slugcomment{Accepted by ApJ}
 
\lefthead{}
\righthead{}
 
\begin{document}
 
\title{ Critical Protoplanetary Core Masses in Protoplanetary Disks
and the Formation of Short--Period Giant Planets}
 
\author{John C. B. Papaloizou\altaffilmark{1,2} and Caroline
Terquem\altaffilmark{2,3,4}}
 
\altaffiltext{1}{Astronomy Unit, School of Mathematical Sciences,
Queen Mary~\& Westfield College, Mile End Road, London E1~4NS, UK --
J.C.B.Papaloizou@qmw.ac.uk}

\altaffiltext{2}{Isaac Newton Institute for Mathematical Sciences,
University of Cambridge, 20 Clarkson Road, Cambridge CB3~0EH, UK}

\altaffiltext{3}{UCO/Lick Observatory, University of California,
Santa~Cruz, CA~95064, USA -- ct@ucolick.org}

\altaffiltext{4}{On leave from: Laboratoire d'Astrophysique,
Observatoire de Grenoble, Universit\'e Joseph Fourier/CNRS, BP~53,
38041 Grenoble Cedex 9, France}
 
\begin{abstract}

We study a solid protoplanetary core undergoing radial migration in a
protoplanetary disk.  We consider cores in the mass range $\sim
1-10$~M$_{\oplus}$ embedded in a gaseous protoplanetary disk at
different radial locations.

We suppose the core luminosity is generated as a result of
planetesimal accretion and calculate the structure of the gaseous
envelope assuming hydrostatic and thermal equilibrium.  This is a good
approximation during the early growth of the core while its mass is
less than the critical value, $M_{crit}$, above which such static
solutions can no longer be obtained and rapid gas accretion
begins. The critical value corresponds to the crossover mass above
which rapid gas accretion begins in time dependent calculations.

We model the structure and evolution of the protoplanetary nebula as
an accretion disk with constant $\alpha$.  We present analytic fits
for the steady state relation between disk surface density and mass
accretion rate as a function of radius.

We calculate $M_{crit}$ as a function of radial location, gas
accretion rate through the disk, and planetesimal accretion rate onto
the core.  For a fixed planetesimal accretion rate, $M_{crit}$ is
found to increase inwards.  On the other hand it decreases with the
planetesimal accretion rate and hence the core luminosity.

We consider the planetesimal accretion rate onto cores migrating
inwards in a characteristic time $\sim 10^3-10^5$~yr at 1~AU as
indicated by recent theoretical calculations.  We find that the
accretion rate is expected to be sufficient to prevent the attainment
of $M_{crit}$ during the migration process if the core starts off
significantly below it.  Only at those small radii where local
conditions are such that dust, and accordingly planetesimals, no
longer exist can $M_{crit}$ be attained.

At small radii, the runaway gas accretion phase may become longer than
the disk lifetime if the mass of the core is too small.  However,
within the context of our disk models, and if it is supposed that some
process halts the migration, massive cores can be built--up through
the merger of additional incoming cores on a timescale shorter than
for {\it in situ \/} formation.  A rapid gas accretion phase may thus
begin without an earlier prolonged phase in which planetesimal
accretion occurs at a reduced rate because of feeding zone depletion
in the neighborhood of a fixed orbit.

Accordingly, we suggest that giant planets may begin to form through
the above processes early in the life of the protostellar disk at small
radii, on a timescale that may be significantly shorter than that
derived for {\it in situ \/} formation.

\end{abstract}


\section{Introduction}
\label{sec:intro}

The hypothesis that the planets in the solar system were formed in a
flattened differentially rotating gaseous disk was originally proposed
by Kant (1755) and Laplace (1796).  Since then, the presence of disks
around low--mass young stellar objects has been inferred from their
infrared excess (Adams, Lada~\& Shu 1987).  Recently they have also
been imaged directly with the {\it Hubble Space Telescope \/}
(McCaughrean~\& O'Dell 1996; Burrows et al. 1996; McCaughrean et
al. 1998; Krist et al. 1998; Stapelfeldt et al. 1998).  Surveys in the
Orion nebula (Stauffer et al. 1994) and the Taurus--Auriga dark clouds
(Beckwith et al. 1990) indicate that these disks are common,
apparently surrounding between 25 and 75\% of the young stellar
objects.  Their infrared emission may be produced by the gravitational
potential energy liberated by matter flowing inwards at a rate
$\dot{M} \sim 10^{-8 \pm 1}$~M$_{\sun}$~yr$^{-1}$ (Hartmann et
al. 1998).  The nonobservation of disks around older T~Tauri stars
together with these values of $\dot M$ suggest a disk lifetime of
between $10^6$ and $10^7$~yr (Strom, Edwards \& Skrutskie 1993).
Masses between $10^{-3}$ and $10^{-1}$~M$_{\odot}$ and dimensions in
the range 10--100~AU have been estimated (Beckwith~\& Sargent 1996).

Most theoretical protostellar disk models have relied on the
$\alpha$--parametrization proposed by Shakura~\& Sunyaev (1973).  In
this context, the disk anomalous turbulent viscosity, which enables
angular momentum to be transported outwards and therefore matter to
flow inwards and be ultimately accreted by the central star, is
assumed to give rise to a stress tensor which is simply proportional
to the gas pressure.  So far, only MHD instabilities (Balbus~\& Hawley
1991) have been shown to be able to produce and sustain turbulence in
accretion disks, and they do lead to an $\alpha$--type disk (Balbus~\&
Hawley 1998).  However, because these instabilities develop only in an
adequately ionized fluid, they may not operate everywhere in
protostellar disks (Gammie 1996).  Therefore it is likely that the
parameter $\alpha$, to which the viscosity is simply related, is not
constant through these disks.  It may even be that only parts of these
disks can be described using this $\alpha$ prescription for the
viscosity.  However, we may still learn about disks from these models
in the same way as we learned about stars from simple polytropic
models. Therefore, for the purpose of considering planet formation, as
we are interested in here, we will use such models.

Planets are believed to form out of protostellar disks by either
gravitational instability (Kuiper 1951; Cameron 1978; Boss 1998) or by
a process of growth through planetesimal accumulation followed, in the
giant planet case, by gas accretion (Safronov 1969; Wetherill~\&
Stewart 1989; Perri~\& Cameron 1974; Mizuno 1980; Bodenheimer~\&
Pollack 1986).  The first mechanism is expected to produce
preferentially massive objects in the outer parts of the disk, if
anything.  In this paper we will study planetary formation within the
context of the second mechanism, which is commonly accepted as the
most likely process by which planets form in at least the inner ten
astronomical units of protostellar disks.  We note however that
important issues related to this model still remain to be resolved
(see Lissauer 1993 for a review).

Up to now, planetary formation has been studied at a given location in
a disk.  Most work has concentrated on orbital distances
corresponding to the neighborhood of Jupiter.  More recently,
prompted by the detection of planets orbiting at short distances from
their host star, {\it in situ \/} formation of giant planets at these
locations has also been considered (Ward 1997a; Bodenheimer 1998;
Bodenheimer, Hubickyj~\& Lissauer 1998).  

However, the importance of orbital migration as recently indicated by
both observations (see Marcy and Butler 1998 and references therein)
and theory (see Lin~\& Papaloizou 1993 and references therein; Ward
1997b) suggests that planets may not form at a fixed location in the
disk, but more likely grow while migrating through the nebula.  It is
the purpose of this paper to investigate the effect of migration on
planetary formation.

In \S~\ref{sec:disk_models} we construct steady state $\alpha$--disk
models for $\alpha=10^{-2}$ and $\alpha =10^{-3}$.  A range of gas
accretion rates, $\dot{M}$, varying between $10^{-6}$ and
$10^{-9}$~M$_{\odot}$~yr$^{-1}$ are considered.  The steady state
assumption is reasonable in the inner regions, below 5~AU from the
central star, where the local viscous timescale is short, typically on
the order of $10^4$~yr for $\dot{M}=10^{-7}$~M$_{\odot}$~yr$^{-1}$.
We present analytic fits for the steady state relation between disk
surface density and mass accretion rate as a function of disk
location.  These fits can be used to solve the diffusion equation
which governs the disk evolution.

In \S~\ref{sec:envelope} we describe the construction of the
protoplanet models based on a solid core with gaseous envelope.  We
review the theory of giant planet formation in \S~\ref{sec:background}
and give the equations governing a protoplanet atmosphere in
\S~\ref{sec:envelope_eq}.  In \S~\ref{sec:envelope_calc} we give the
results of numerical calculations for the critical core mass above
which the atmosphere cannot remain in hydrostatic and thermal
equilibrium but must evolve, with the protoplanet entering a rapid gas
accretion phase.  These calculations are done at different locations
in our disk models.  For a given planetesimal accretion rate which
supplies the core luminosity necessary to support the envelope,
critical core masses are found to increase significantly by factors of
2--3 between 5~AU and 0.05~AU, at which point local conditions may not
enable planetesimals to exist.

In \S~\ref{sec:migration} we consider migration of the protoplanetary
cores which, according to recent estimates of the effects of tidal
interactions with the disk by Ward (1997b), may occur on a timescale
between $ 10^4$~yr and $ 10^5$~yr for a core mass of several earth
masses at 5~AU. This is also comparable to the proposed formation
time.  We perform simulations which indicate that such a migrating
protoplanet is likely to accrete 24\% or more of the planetesimals
initially interior to it.  This accretion is likely to maintain the
core luminosity such that attainment of the critical mass does not
occur until it reaches small radii $\sim 0.1$~AU where planetesimals
no longer exist.

In \S~\ref{sec:discussion} we discuss our results in the context of
short--period giant planets.  We point out that the processes
investigated in this paper are likely to result in a giant planet
orbiting at small radii on a timescale significantly shorter than that
derived for {\it in situ \/} formation.  Finally in
\S~\ref{sec:summary} we summarize our results.

\section{Disk Models}
\label{sec:disk_models}

\subsection{Vertical Structure}

\subsubsection{Basic Equations}
\label{sec:equations}

Here we consider the equations governing the disk vertical structure
in the thin--disk approximation.  Using cylindrical coordinates
$(r,\varphi, z)$ based on the central star and such that the $z=0$
plane corresponds to the disk midplane, we adopt the equation of
vertical hydrostatic equilibrium in the form:

\begin{equation}
\frac{1}{\rho} \frac{\partial P}{\partial z} = - \Omega^2 z ,
\label{dPdz}
\end{equation}

\noindent together with the energy equation, which states that the
rate of energy removal by radiation is locally balanced by the rate of
energy production by viscous dissipation:

\begin{equation}
\frac{\partial F}{\partial z} = \frac{9}{4} \rho \nu \Omega^2 ,
\label{dFdz1}
\end{equation}

\noindent where $F$ is the radiative flux of energy through a
surface of constant $z$  which is given by:

\begin{equation}
F = \frac{- 16 \sigma T^3}{3 \kappa \rho}
\frac{\partial T}{\partial z} .
\label{dTdz}
\end{equation}

\noindent Here $\rho$ is the mass density per unit volume, $P$ is the
pressure, $T$ is the temperature, $\Omega$ is the angular velocity,
$\nu$ is the kinematic viscosity, $\kappa$ is the opacity, which in
general depends on both $\rho$ and $T$, and $\sigma$ is the
Stefan--Boltzmann constant.  We consider thin disks which are in
Keplerian rotation around a star of mass $M_{\ast}$, so that $\Omega^2
= GM_{\ast}/r^3$, $G$ being the gravitational constant.   In
writing equation~(\ref{dTdz}), we have assumed that the disk at the
radius considered is optically thick.  However, when the disk is
optically thin, i.e. when $\kappa \rho$ integrated over the disk
thickness is small compared to unity, the temperature gradient given
by equation~(\ref{dTdz}) is small, so that the results we get are
consistent in that case also.

To close the system of equations, we relate $P$, $\rho$ and $T$
through the equation of state of an ideal gas:

\begin{equation}
P = \frac{\rho k T}{\mu m_H} ,
\label{state}
\end{equation}

\noindent where $k$ is the Boltzmann constant, $\mu$ is the mean
molecular weight and $m_H$ is the mass of the hydrogen atom. Here we
shall limit our calculations to temperatures lower than 4,000~K, so
that, at the densities of interest, hydrogen is in molecular form.
Since the main component of protostellar disks is hydrogen, it is a
reasonable approximation to take $\mu=2$.  Tests using a more
sophisticated equation of state such as that of Chabrier
et. al. (1992) indicate only minor differences in the results for the
range of temperatures and densities considered.  Similarly transport
of energy by convection can be neglected here (see also
Lin~\&~Papaloizou~1985).

We adopt the $\alpha$--parametrization of Shakura \& Sunyaev (1973),
so that the kinematic viscosity is written $\nu=\alpha c_s^2/\Omega$,
where $c_s$ is the isothermal sound speed ($c_s^2=P/\rho$).  Although
in general $\alpha$ may be a function of both $r$ and $z$, we shall
limit our calculations presented below to cases with constant $\alpha$
(see discussion in \S~\ref{sec:intro}).  With this formalism,
equation~(\ref{dFdz1}) becomes:

\begin{equation}
\frac{\partial F}{\partial z} = \frac{9}{4} \alpha \Omega P .
\label{dFdz}
\end{equation}

\subsubsection{Boundary Conditions}
\label{sec:boundary}

We have to solve three first order ordinary differential equations for
the three variables $F$, $P$ (or equivalently $\rho$), and $T$
as a function of $z$ at a given radius $r$.  Accordingly, we need
three boundary conditions at each $r$.  We denote with a subscript $s$
values at the disk surface.
                            
A boundary condition is obtained by integrating equation~(\ref{dFdz1})
over $z$ between $-H$ and $H$ which define the lower and upper
boundary of the disk, respectively.  Since by symmetry $F(z=0)=0$,
this gives:

\begin{equation}
F_s = \frac{3}{8 \pi} \dot{M}_{st} \Omega^2 ,
\label{Fs}
\end{equation}

\noindent where we have defined $\dot{M}_{st} = 3 \pi
\langle{\nu}\rangle \Sigma$ with $\Sigma = \int_{-H}^H \rho dz$ being
the disk surface mass density and $\langle \nu \rangle = \int_{-H}^H
\rho \nu dz/ \Sigma$ being the vertically averaged viscosity.  If the
disk were in a steady state, $\dot{M}_{st}$ would not vary with $r$
and would be the constant accretion rate through the disk.  In general
however, the disk is not in a steady state (but see
\S~\ref{sec:evolution}) such that this quantity does depend on $r$ and
the disk undergoes time dependent evolution.

Another boundary condition is obtained by integrating
equation~(\ref{dPdz}) over $z$ between $H$ and infinity.  A detailed
derivation of this condition is presented in Appendix~\ref{appendixA}.
Here we just give the result:

\begin{equation}
P_s = \frac{\Omega^2 H \tau_{ab}}{\kappa_s} ,
\label{Ps}
\end{equation}

\noindent where $\tau_{ab}$ is the optical depth above the disk.  This
condition is familiar in stellar structure, where $\Omega^2 H$ would
be replaced by the acceleration of gravity at the stellar surface
(e.g. Schwarzschild 1958).  Since we have defined the disk surface
such that the atmosphere above the disk is isothermal, we have to take
$\tau_{ab} \ll 1$.  Providing this is satisfied, the results do not
depend on the value of $\tau_{ab}$ we choose (see
\S~\ref{sec:calcul}).

A third and final boundary condition is given by the expression of the
surface temperature (see Appendix~\ref{appendixA} for a detailed
derivation of this expression):

\begin{equation}
2 \sigma \left( T_s^4 - T_b^4 \right) - \frac{9 \alpha k T_s \Omega}{8
\mu m_H \kappa_s} - \frac{3}{8 \pi} \dot{M}_{st} \Omega^2 = 0 .
\label{Ts}
\end{equation}

\noindent Here the disk is assumed immersed in a medium with
background temperature $T_b$.  The surface opacity $\kappa_s$ in
general depends on both $T_s$ and $\rho_s$ and we have used
$c_s^2=kT/(\mu m_H)$.  The boundary condition~(\ref{Ts}) is the same
as that used by Levermore \& Pomraning (1981) in the Eddington
approximation (their eq.~[56] with $\gamma=1/2$).  In the simple case
when $T_b=0$ and the surface dissipation term involving $\alpha$ is
set to zero, with $\dot{M}_{st}$ being retained, it simply relates the
disk surface temperature to the emergent radiation flux.

\subsubsection{Model Calculations}
\label{sec:calcul}

At a given radius $r$ and for a given value of the parameters
$\dot{M}_{st}$ and $\alpha$, we solve equations~(\ref{dPdz}),
(\ref{dTdz}) and~(\ref{dFdz}) with the boundary conditions~(\ref{Fs}),
(\ref{Ps}) and~(\ref{Ts}) to find the dependence of the state
variables on $z$.  The opacity is taken from Bell~\& Lin (1994). This
has contributions from dust grains, molecules, atoms and ions.  It is
written in the form $\kappa=\kappa_i \rho^a T^b$ where $\kappa_i$, $a$
and $b$ vary with temperature.

\noindent The equations are integrated using a fifth--order
Runge--Kutta method with adaptive step length (Press et al. 1992).
For a specified $\dot{M}_{st}$ and $\alpha$, we first calculate the
surface flux $F_s$ and temperature $T_s$ from
equations~(\ref{Fs}) and~(\ref{Ts}), respectively.  If $T_s$ is
smaller than about 1,000~K, the opacity $\kappa_s$ at the disk surface
does not depend on $\rho_s$.  For larger values of $T_s$, it turns out
that the second term in equation~(\ref{Ts}), which contains
$\kappa_s$, is negligible compared to the third term.  Therefore $T_s$
can always be calculated independently of $\rho_s$ (or equivalently
$P_s$).  We then determine the value of $H,$ the vertical height of
the disk surface, iteratively.  Starting from an estimated value of
$H,$ we calculate $P_s$ and integrate the equations from $H$ down to
the midplane $z=0$.  The condition that $F=0$ at $z=0$ will not
in general be satisfied.  An iteration procedure is then used to
adjust the value of $H$ until $F=0$ at $z=0$ to a specified
accuracy.

An important point to note is that as well as finding the disk
structure, we also determine the surface density $\Sigma$ for a given
$\dot{M}_{st} =3\pi\langle \nu \rangle \Sigma$.  In this way, a
relation between $\langle \nu \rangle$ and $\Sigma$ is derived.

In the calculations presented here, we have taken $M_{\ast} =
1$~M$_{\sun}$, the optical depth of the atmosphere above the disk
surface $\tau_{ab}=10^{-2}$ and a background temperature $T_b=10$~K.
In the optically thick regions of the disk, the value of $H$ is
independent of the value of $\tau_{ab}$ we choose.  However, this is
not the case in optically thin regions where we find that, as
expected, the smaller $\tau_{ab},$ the larger $H$.  However, this
dependence of $H$ on $\tau_{ab}$ has no physical significance, since
the surface mass density, the optical thickness through the disk and
the midplane temperature hardly vary with $\tau_{ab}$.  This is
because the mass is concentrated towards the disk midplane in a layer
with thickness independent of $\tau_{ab}.$

\subsection{Time Dependent Evolution and Quasi--Steady States}
\label{sec:evolution}

In general an accretion disk is not in a steady state but undergoes
time dependent evolution.  The global evolution of the disk is
governed by the well known diffusion equation for the surface density
which can be written in the form (see Lynden--Bell~\& Pringle 1974,
Papaloizou~\& Lin 1995 and references therein):

\begin{equation}
{\partial \Sigma \over \partial t} = {3 \over r}{\partial \over
\partial r} \left[ r^{1/2} {\partial \over \partial r} \left( \Sigma
\langle \nu \rangle r^{1/2}\right) \right]
\label{aa4} .
\end{equation}         

\noindent The characteristic diffusion time scale at radius $r$ is
then:

\begin{equation}
t_{\nu} = {r^2\over 3\langle \nu \rangle} \sim \frac{1}{3 \alpha}
\left( {r \over H} \right)^2 \Omega^{-1} .
\end{equation}
 
\noindent For disks with approximately constant aspect ratio $H/r,$ as
applies to the models considered here, $t_{\nu}$ scales as the local
orbital period.  One thus expects that the inner regions relax
relatively quickly to a quasi--steady state which adjusts its
accretion rate according to the more slowly evolving outer parts (see
Lynden--Bell~\& Pringle 1974 and Lin~\& Papaloizou 1985).  For
estimated sizes of protostellar disks of about 50~AU (Beckwith~\&
Sargent 1996), the evolutionary timescale associated with the outer
parts is about 30 times longer than that associated with the inner
parts with $r < 5$~AU, which we consider here in the context of
planetary formation.  Thus these inner regions are expected to be in a
quasi--steady state through most of the disk lifetime.  We have
verified that this is the case by considering solutions of
equation~(\ref{aa4}).

In order to investigate these solutions and for other purposes we
found it convenient to utilize analytic piece--wise power law fits to
the $\langle \nu \rangle$--$\Sigma$ relation derived above (see
\S~\ref{sec:calcul}).  Details of these fits are given in
Appendix~\ref{appendixB}.  In Figures~\ref{fig1}a--b we plot both the
curves $\dot{M}_{st} \left ( \Sigma \right)$ obtained from the
vertical structure integrations and those obtained from the
piece--wise power law fits.  Figures~\ref{fig1}a and~\ref{fig1}b are
for $\alpha=10^{-2}$ and $\alpha=10^{-3}$, respectively.  In each case
the radius varies between 0.01 and 100~AU, and the calculations are
limited to temperatures lower than 4,000~K (see
\S~\ref{sec:equations}).  The average errors are 18\% and 13\% for
$\alpha= 10^{-3}$ and $10^{-2}$, respectively.  Thus the fits give an
adequate approximation.

\subsection{Steady State Models}
\label{sec:steady}

Here, we present solutions corresponding to steady state accretion
disks.  In Figures~\ref{fig1car}a--c and ~\ref{fig2}a--c, we plot
$H/r$, $\Sigma$ and the midplane temperature $T_m$ versus $r$ for
$\dot{M}_{st}$ between $10^{-9}$ and $10^{-6}$~M$_{\odot}$~yr$^{-1}$
(assuming this quantity is the same at all radii, i.e. the disk is in
a steady state) for $\alpha=10^{-2}$ and $\alpha=10^{-3}$,
respectively.

An inspection of Figures~\ref{fig1car}a and~\ref{fig2}a indicates that
the outer parts of the disk are shielded from the radiation of the
central star by the inner parts, apart possibly from the very outer
parts, which are optically thin anyway and therefore do not reprocess
any radiation.  This is in agreement with the results of Lin~\&
Papaloizou (1980) and Bell et al. (1997).  For $\alpha=10^{-3}$, the
radius beyond which the disk is not illuminated by the central star
varies from 0.2~AU to about 3~AU when $\dot{M}_{st}$ goes from
$10^{-9}$ to $10^{-6}$~M$_{\odot}$~yr$^{-1}$.  These values of the
radius move to 0.1 and 2~AU when $\alpha=10^{-2}$.  Since reprocessing
of the stellar radiation by the disk is not an important heating
factor below these radii, this process will in general not be
important in these models of protostellar disks.  We note that this
result is independent of the value of $\tau_{ab}$ we have
taken. Indeed, as we pointed out above, only the thickness of the
optically thin parts of the disk gets larger when $\tau_{ab}$ is
decreased.

However, there are some indications that disks such as HH30 may be
flared (Burrows et al. 1996).  In this context, Chiang~\& Goldreich
(1997) have considered a model based on reprocessing in which the dust
and gas are at different temperatures. However, as they pointed out,
some issues regarding this model remain to be resolved.  In any case,
it is possible that a multiplicity of solutions exists when
reprocessing is taken into account, with it being important for cases
in which the disk is flared and unimportant when it is not, such as
maybe HK~Tau (Stapelfeldt et al. 1998, Koresko 1998).

The values of $H/r$, $\Sigma$ and $T_m$ we get are similar to those
obtained by Lin~\& Papaloizou (1980), who adopted a prescription for
viscosity based on convection, and Bell et al. (1997).  Since $H$ is
measured from the disk midplane to the surface such that $\tau_{ab}$
is small, it is larger than what would be obtained if a value of 2/3
were adopted for $\tau_{ab}$, as is usually the case.  However, this
does not affect other physical quantities.  We also recall that $H,$ as
defined here, is about 2--3 times larger than $c_s/\Omega,$ with $c_s$
being the midplane sound speed, which is commonly used to define the
disk semithickness.

In this paper we shall consider the migration of protoplanetary cores
from $\sim 1$--5~AU, where they are supposed to form under conditions
where ice exists, down to the disk inner radii (see
\S~\ref{sec:migration}).  It is therefore of interest to estimate the
mass of planetesimals contained inside the orbit of a core when it
forms, since this can potentially be accreted by the core during its
migration. From Figures~\ref{fig1car}b and~\ref{fig2}b, we estimate
the mass of planetesimals $M_p(r)$ contained within a radius $r$ using
$M_p(r) = 10^{-2} \times \pi r^2 \Sigma(r)$ for $r=1$~AU and $r=5$~AU.
Here we have assumed a gas to dust ratio of 100.  The values of $M_p$
corresponding to $\alpha=10^{-2}$ and $10^{-3}$ and $\dot{M} =
10^{-6}$ and $10^{-7}$~M$_{\sun}$~yr$^{-1}$ are listed in
Table~\ref{tab1}.  We have checked that the disk with $\alpha=10^{-3}$
and $\dot{M} = 10^{-6}$~M$_{\sun}$~yr$^{-1}$, although relatively
massive, is gravitationally stable locally. Namely the Toomre
parameter $Q=(M_{\ast}/M(r))(H/r),$ with $M(r)=\pi r^2 \Sigma(r)$, is
larger than unity.

It is also of interest, in relation to the possibility of giant
planets being located at small radii, to estimate the mass of gas
contained within a radius of 0.1~AU.  Figure~\ref{fig2}b indicates
that, when $\alpha= 10^{-3}$, this mass is about 0.3 Jupiter mass for
${\dot M} > 10^{-7}$~M$_{\odot}$~yr$^{-1}$.  For $\alpha =10^{-2}$,
there is a similar mass of gas for ${\dot M} >
10^{-6}$~M$_{\odot}$~yr$^{-1}$ (see Figure~\ref{fig1car}b).  For
typical mass throughput of about $10^{-2}$--$10^{-1}$~M$_{\odot},$ the
lifetime of such a state can range between $10^4$ and $10^6$~yr.
Supposing the disk to be terminated at some small inner radius, this
suggests that, if a suitable core can migrate there, it could accrete
enough gas to become a giant planet within the disk lifetime.  We note
however that the conditions for that to happen are marginal even in
the early stages of the life of the disk when ${\dot M} >
10^{-6}-10^{-7}$~M$_{\odot}$~yr$^{-1}.$

At later stages, when $\dot M \sim 10^{-8}$~M$_{\odot}$~yr$^{-1},$ the
models resemble conditions expected to apply to the minimum mass solar
nebula with $\Sigma \sim 200$~g~cm$^{-2}$ at 5~AU if $\alpha=10^{-2}$.
Under these conditions, the mass of gas at $r < 0.1$~AU is between 1
and 9~M$_{\oplus}$ for $\alpha$ between $10^{-2}$ and $10^{-3}$.

Lin et al. (1996) suppose that the inner disk is terminated by a
magnetospheric cavity. In this case migration might be supposed to
cease if the core is sufficiently far inside it.  But gas accretion is
likely to be very much reduced in this case.

However, the protoplanet may be able to accrete more gas than the mass
estimated above if circumstances were such that the core migration is
halted at some small radius before the disk is terminated.  We note
that Ward (1986) finds the direction of type~I migration is
insensitive to the disk surface density profile but that it could
reverse from inwards to outwards if the disk midplane temperature
decreased inwards faster than approximately linearly. Such a condition
would not be expected in the disk models considered here.

On the other hand, conditions may be very different if interaction
with a stellar magnetic field becomes important. It is expected that
this happens when magnetic and viscous torques become comparable
(e.g. Ghosh \& Lamb 1979). In this region there may be open field
lines connected to the disk with an outflowing wind (Paatz \&
Camenzind 1996).  Such a wind may provide an additional angular
momentum and energy loss mechanism for the disk material (Papaloizou
\& Lin 1995).  The inner regions could then be cooler than expected
from the constant $\alpha$ models considered here leading to a
reversal of type~I migration.  A faster gas inflow rate may also
prevent the onset of type~II migration. Thus, although details are
unclear, continued accretion of inflowing disk gas by a protoplanetary
core that has stopped migrating in the inner disk may be possible.

\section{Protoplanetary Core Growth and Equilibrium Envelope} 
\label{sec:envelope}

\subsection{Background: Formation of Giant Planets}
\label{sec:background}

The solid cores of giant  planets are believed to be formed via solid body
accretion of km--sized planetesimals, which themselves are produced as
a result of the sedimentation and collisional growth of dust grains in
the protoplanetary disk (see Lissauer 1993 and references therein).
Once the solid core becomes massive enough to gravitationally bind the
gas in which it is embedded (typically at a tenth of an earth mass), a
gaseous envelope begins to form around the core.

The build--up of the atmosphere has first been considered in the
context of the so--called 'core instability' model by Perri~\& Cameron
(1974) and Mizuno (1980; see also Stevenson 1982 and Wuchterl 1995).
In this model, the solid core grows in mass along with the atmosphere
in quasi--static and thermal equilibrium until the core reaches the
so--called 'critical core mass' above which no equilibrium solution
can be found for the atmosphere.  As long as the core mass is smaller
than the critical core mass, the energy  radiated from the envelope into
the surrounding nebula is compensated for by the gravitational energy
which the planetesimals entering the atmosphere release when they
collide with the surface of the core.  During this phase of the
evolution, both the core and the atmosphere grow in mass relatively
slowly.  By the time the core mass reaches the critical core mass, the
atmosphere has grown massive enough so that its energy losses can no
longer be compensated for by the accretion of planetesimals alone.  At
that point the envelope has to contract gravitationally to supply more
energy.  This is a runaway process, leading to the very rapid
accretion of gas onto the protoplanet and to the formation of giant
planets such as Jupiter.  In  earlier studies it was assumed that
this rapid evolution was a dynamical collapse, hence the designation
'core instability' for this model.  

Further time--dependent numerical calculations of protoplanetary
evolution by Bodenheimer~\& Pollack (1986) support this model,
although they show that the core mass beyond which runaway gas
accretion occurs, which is referred to as the 'crossover mass', is
slightly larger than the critical core mass, and that the very rapid
gravitational contraction of the envelope is not a dynamical collapse.
The designation 'crossover mass' comes from the fact that rapid
contraction of the atmosphere occurs when the mass of the atmosphere
is comparable to that of the core.  Once the crossover mass is
reached, the core no longer grows significantly.

More recent simulations by Pollack et al. (1996) show that the
evolution of a protoplanet is governed by three distinct phases.
During phase~1, runaway planetesimal accretion occurs which leads to
the depletion of the feeding zone of the protoplanet.  At this point,
when phase~2 begins, the atmosphere is massive enough that the
location of its outer boundary is determined by both the mass of gas
and planetesimals.  As more gas is accreted, this outer radius moves
out, so that the feeding zone is increased and more planetesimals can
be captured, which in turn enables more gas to enter the atmosphere.
The protoplanet grows in this way until the core reaches the crossover
mass, at which point runaway gas accretion occurs and phase~3 begins.
The timescale for planet formation is determined almost entirely by
phase~2, and is found to be a few million years at 5~AU. For typical
disk models, this is comparable to the disk lifetime.  Note however
that isolation of the protoplanetary core, as it occurs at the end of
phase~1, may be prevented under some circumstances by tidal
interaction with the surrounding gaseous disk (Ward~\& Hahn 1995).

Conditions appropriate to Jupiter's present orbital radius are
normally considered and then the critical or crossover mass is found
to be around 15~M$_{\oplus}$.  This is consistent with models of
Jupiter which indicate that it has a solid core of about
5--15~M$_{\oplus}$ (Podolak et al. 1993).

We note that although phase~3 is relatively rapid compared to phase~2
for conditions appropriate to Jupiter's present location, it may
become longer when the luminosity provided by the accretion of
planetesimals, and hence the critical core mass, is reduced (see
Pollack et al. 1996 and \S~\ref{sec:discussion}).  The designation
'runaway' or 'rapid' gas accretion may then become confusing.

The similarity between the critical and crossover masses is due to the
fact that, although there is some liberation of gravitational energy
as the atmosphere grows in mass together with the core, the effect is
small as long as the atmospheric mass is small compared to that of the
core.  Consequently the hydrostatic and thermal equilibrium
approximation for the atmosphere is a good one for core masses smaller
than the critical value.  Therefore we use this approximation here and
investigate how the critical core mass varies with location and
physical conditions in the protoplanetary disk.

\subsection{Basic Equations Governing a Protoplanetary Envelope}
\label{sec:envelope_eq}

Let $R$ be the spherical polar radius in a frame with origin at the
center of the protoplanet's core.  We assume that we can model the
protoplanet as a spherically symmetric nonrotating object.  We also
assume that it is in hydrostatic and thermal equilibrium.  The
equation of hydrostatic equilibrium is then:

\begin{equation}
\frac{dP}{d R} = - g \rho ,
\label{dpdvarpi}
\end{equation}

\noindent where $g=G M / R^2$ is the acceleration due to gravity,
$M(R)$ being the mass contained in the sphere of radius $R$
(this includes the core mass if $R$ is larger than the core
radius).  We also have the definition of density:

\begin{equation}
\frac{dM}{d R} = 4 \pi R^2 \rho.
\label{dmdvarpi}
\end{equation}

At the high densities that occur at the base of a protoplanetary
envelope, the gas cannot be considered to be ideal.  Thus we adopt the
equation of state for a hydrogen and helium mixture given by Chabrier
et al. (1992).  We adopt the mass fractions of hydrogen and helium to
be 0.7 and 0.28, respectively.  We also use the standard equation of
radiative transport in the form:

\begin{equation}
\frac{dT}{d R} = \frac{-3 \kappa \rho}{16 \sigma
T^3} \frac{L}{4 \pi R^2} .
\label{dtdvarpi}
\end{equation}

\noindent Here $L$ is the  radiative luminosity.  Denoting
the radiative and adiabatic temperature gradients by $\nabla_{rad}$
and $\nabla_{ad}$, respectively, we have:

\begin{equation}
\nabla_{rad} = \left( \frac{\partial \ln T}{\partial \ln P}
\right)_{rad} = \frac{3 \kappa L_{core} P}{64 \pi
\sigma G M T^4} ,
\label{dTdr_rad}
\end{equation}

\noindent and

\begin{equation}
\nabla_{ad} = \left( \frac{\partial \ln T}{\partial \ln P} \right)_s ,
\end{equation}

\noindent with the subscript $s$ denoting evaluation at constant
entropy.

We assume that the only energy source comes from the accretion of
planetesimals onto the core which, as a result, outputs a total core
luminosity $L_{core}$, given by:

\begin{equation}
L_{core} = \frac{ G M_{core} \dot{M}_{core}}{ r_{core}} .
\end{equation}

\noindent Here $M_{core}$ and $r_{core}$ are respectively the mass and
the radius of the core, and $\dot{M}_{core}$ is the planetesimal
accretion rate.  The luminosity $L_{core}$ is supplied by the
gravitational energy which the planetesimals entering the planet
atmosphere release near the surface of the core (see, e.g., Mizuno
1980; Bodenheimer \& Pollack 1986).

\noindent If $\nabla_{rad} < \nabla_{ad}$, there is stability to
convection and thus all the energy is transported by radiation,
i.e. $L=L_{core}$.  When $\nabla_{rad} > \nabla_{ad}$, there is
instability to convection. Then, part of the energy is transported by
convection, and $L_{core}= L+L_{conv}$, where $L_{conv}$ is the
luminosity associated with convection.  We use mixing length theory to
evaluate $L_{conv}$ (Cox~\& Giuli 1968).  Then

\begin{eqnarray}
L_{conv} & = & \pi R^2 C_p \Lambda_{ml}^2 \left[ \left( \frac{\partial
T}{\partial R} \right)_s - \left( \frac{\partial T}{\partial R}
\right) \right]^{3/2}
\nonumber \\
& \times & 
\sqrt{ \frac{1}{2} \rho g \left| \left(
\frac{\partial \rho}{\partial T} \right)_P \right| } ,
\end{eqnarray}

\noindent where $\Lambda_{ml}=|\alpha_{ml}P/(dP/d R)|$ is the mixing
length, $\alpha_{ml}$ being a constant of order unity, $\left(
\partial T/\partial R \right)_s = \nabla_{ad} T \left( d \ln P / d R
\right)$, and the subscript $P$ means that the derivative has to be
evaluated for a constant pressure.  All the required thermodynamic
parameters are given by Chabrier et al. (1992).  In the numerical
calculations presented below we fix $\alpha_{ml}=1$.

\subsubsection{Inner Boundary}

We suppose that the planet core has a uniform mass density
$\rho_{core}$.  The composition of the planetesimals and the high
temperatures and pressures at the surface of the core suggest
$\rho_{core}=3.2$~g~cm$^{-3}$ (see Bodenheimer~\& Pollack 1986 and
Pollack et al. 1996), which is the value we will adopt throughout.
The core radius, which is the inner boundary of the atmosphere, is
then given by:

\begin{equation}
r_{core} = \left( \frac{3 M_{core}}{4 \pi \rho_{core}} \right)^{1/3}.
\end{equation}
At $R=r_{core}$ the total mass is equal to $M_{core}.$

\subsubsection{Outer Boundary}

We take the outer boundary of the atmosphere to be at the Roche lobe
radius $r_L$ of the protoplanet. Thus:

\begin{equation}
r_L = \frac{2}{3} \left( \frac{M_{pl}}{3 M_{\ast}} \right)^{1/3} r ,
\end{equation}

\noindent where $M_{pl} = M_{core} + M_{atm}$ is the planet mass,
$M_{atm}$ being the mass of the atmosphere, and $r$ is the orbital
radius of the protoplanet in the disk.

To avoid confusion, we will denote the disk midplane temperature,
pressure and mass density at the distance $r$ from the central star by
$T_m,$ $P_m$ and $\rho_m$, respectively.

\noindent At $R=r_L,$ the mass is equal to $M_{pl}$, the pressure
is equal to $P_m$ and the temperature is given by

\begin{equation}
T = \left( T_m^4 + \frac{ 3\tau_L L_{core}}{16 \pi \sigma r_L^2}
\right)^{1/4},
\end{equation}

\noindent where we approximate the additional optical depth above the
protoplanet atmosphere, through which radiation passes, by:

\begin{equation}
\tau_L = \kappa \left( \rho_m, T_m \right) \rho_m r_L .
\end{equation}

\subsection{Calculations}
\label{sec:envelope_calc}

For a particular disk model, at a chosen radius $r,$ for a given core
mass $M_{core}$ and planetesimal accretion rate $\dot{M}_{core}, $ we
solve the equations~(\ref{dpdvarpi}), (\ref{dmdvarpi})
and~(\ref{dtdvarpi}) with the boundary conditions described above to
get the structure of the envelope.  The opacity law adopted is the
same as that for the disk models.  In general, the deep interior of
the envelope becomes convective with the consequence that the value of
the opacity does not matter there.

For a fixed $\dot{M}_{core}$ at a given radius, there is a critical
core mass $M_{crit}$ above which no solution can be found, i.e. there
can be no atmosphere in hydrostatic and thermal equilibrium confined
between the radii $r_{core}$ and $r_L$ around cores with mass larger
than $M_{crit}$, as explained in \S~\ref{sec:background}.  For masses
below $M_{crit}$, there are (at least) two solutions, corresponding to
a low--mass and a high--mass envelope, respectively.

In Figure~\ref{fig4} we plot curves of total protoplanet mass $M_{pl}$
against core mass $M_{core}$ at different radii in a disk with
$\alpha= 10^{-2}$ and ${\dot M}=10^{-7}$~M$_{\sun}$~yr$^{-1}$.  In
each frame, the different curves correspond to planetesimal accretion
rates in the range $10^{-11}$--$10^{-6}$~M$_{\oplus}$~yr$^{-1}$.  The
critical core mass is attained at the point where the curves start to
loop backwards.

When the core first begins to gravitationally bind some gas, the
protoplanet is on the left on the lower branch of these curves.
Assuming $\dot{M}_{core}$ to be constant, as the core and the
atmosphere grow in mass, the protoplanet moves along the lower branch
up to the right, until the core reaches $M_{crit}$.  At that point the
hydrostatic and thermal equilibrium approximation can no longer be
used for the atmosphere, which begins to undergo very rapid
contraction.  Figure~\ref{fig4} indicates that when the core mass
reaches $M_{crit}$, the mass of the atmosphere is comparable to that
of the core, in agreement with Bodenheimer~\& Pollack (1986).  Since
the atmosphere in complete equilibrium is supported by the energy
released by the planetesimals accreted onto the protoplanet, we expect
the critical core mass to decrease as $\dot{M}_{core}$ is reduced.
This is indeed what we observe in Figure~\ref{fig4}.

For $\alpha=10^{-2}$ and $\dot{M}=10^{-7}$~M$_{\sun}$~yr$^{-1}$, the
critical core mass at 5~AU varies between 16.2 and 1~M$_{\oplus}$ as
the planetesimal accretion rate varies between the largest and
smallest value.  The former result is in good agreement with that of
Bodenheimer~\& Pollack (1986).  Note that there is a tendency for the
critical core masses to increase as the radial location moves inwards,
the effect being most marked at small radii.  At 1~AU, the critical
mass varies from 17.5 to 3~M$_{\oplus}$ as the accretion rate varies
between the largest and smallest value, while at 0.05~AU these values
increase still further to 42 and 9~M$_{\oplus}$, respectively.  We
note there has been some debate over the amount of grain opacity which
should be used in these calculations.  However, the results of
Bodenheimer~\& Pollack (1986) indicate that $M_{crit}$ does not depend
sensitively on the grain opacity in the envelope.

In Figure~\ref{fig5} we plot the critical core mass $M_{crit}$ versus
the location $r$ for three different steady disk models.  These models
have $\alpha=10^{-2}$ and $\dot{M} = 10^{-7}$~M$_{\odot}$~yr$^{-1}$,
$\alpha=10^{-2}$ and $\dot{M} = 10^{-8}$~M$_{\odot}$~yr$^{-1}$, and
$\alpha=10^{-3}$ and $\dot{M} = 10^{-8}$~M$_{\odot}$~yr$^{-1}$,
respectively.  Here again, in each frame, the different curves
correspond to planetesimal accretion rates in the range
$10^{-11}$--$10^{-6}$~M$_{\oplus}$~yr$^{-1}$.  Similar qualitative
behavior is found for the three disk models, but the critical core
masses are smaller for the models with $\dot{M} =
10^{-8}$~M$_{\odot}$~yr$^{-1}$, being reduced to 27 and 6~M$_{\oplus}$
at 0.05~AU for the highest and lowest accretion rate, respectively.

These results indicate a relatively weak dependence on disk conditions
except when rather high midplane temperatures $T_m > 1,000$~K are
attained, as in the inner regions.  We indeed find similar values of
$M_{crit}$ for the three different models at radii larger than about
0.15--0.5~AU, where $T_m$ is lower than 1,000~K.  Also, the fact that
$M_{crit}$ is similar for the two models with $\dot{M} =
10^{-8}$~M$_{\odot}$~yr$^{-1}$ indicates that $M_{crit}$ is more
sensitive to the midplane temperature than to the midplane pressure.
These two models have indeed similar $T_m$, whereas $P_m$ varies
significantly from one model to the other.  In the model with $\dot{M}
= 10^{-7}$~M$_{\odot}$~yr$^{-1}$, $T_m$ is significantly larger, hence
the larger $M_{crit}$ at small radii in this case.  These results are
consistent with the fact that $M_{crit}$ depends on the boundary
conditions only when a significant part of the envelope is convective
(Wuchterl 1993), being larger for larger convective envelopes
(Perri~\& Cameron 1974).  In the model with $\alpha=10^{-2}$ and
$\dot{M}=10^{-7}$~M$_{\sun}$~yr$^{-1}$, we indeed find that the inner
70\% in radius of the envelope is convective at 0.05~AU, this value
being reduced to 10\% at 5~AU.  When the envelope is mainly radiative,
it converges rapidly to the radiative zero solution independently of
its outer boundary conditions, so that $M_{crit}$ hardly depends on
the background temperature and pressure (Mizuno 1980).

We note that it is unlikely there are planetesimals at the smallest
radii considered here.  Therefore, although critical core masses for
the same planetesimal accretion rates may be higher there, a lack of
planetesimals may result in a fall in the core luminosity, making the
critical core mass relatively small at these radii.

\section{Protoplanet Migration and Planetesimal Accretion}
\label{sec:migration}

According to current models of planet formation (Safronov 1969;
Wetherill~\& Stewart 1989), planetesimals form as the result of the
coagulation of dust grains.  Further accumulation through binary
collisions then results in the formation of a core of several earth
masses.  After the critical mass is attained, runaway gas accretion
may begin (see \S~\ref{sec:background}).  The core may form on a
timescale of about $10^5$~yr at a distance of about 5~AU as a result
of runaway planetesimals accretion (Lissauer~\& Stewart 1993).

In addition, dynamical tidal interaction of a core of several earth
masses with the surrounding disk matter becomes important, leading to
phenomena such as inward orbital migration and gap formation (Lin~\&
Papaloizou 1979a; Goldreich~\& Tremaine 1980; Lin~\& Papaloizou 1993;
Korycansky \& Pollack 1993; Ward 1997b).  For conditions under which
the tidal interaction with the disk is linear, Ward (1986, 1997b)
estimates inward migration (referred to as type~I) timescales
$t_{mig}= -2r(dr/dt)^{-1}$ of about $2 \times 10^{5} \left( M_{pl}/
{\rm M}_{\oplus} \right)^{-1}$~yr at 5~AU in a protoplanetary disk
similar to the minimum mass solar nebula.  The timescales at other
radii are expected to roughly scale as $\Omega^{-1}$ for the model
disks considered here.  They are also somewhat shorter during the
early phases of disk evolution when the disk is more massive.

However, for core masses of the magnitude we consider, the interaction
may become nonlinear, leading to a reduction in the inward migration
(then referred to as type~II) rate.  To investigate the conditions for
nonlinearity, Korycansky~\& Papaloizou (1996) considered the perturbed
disk flow around an embedded protoplanet assuming the disk viscosity
to be negligible.  They used a shearing--sheet approximation in which
a patch, centered on the planet and corotating with its orbit, is
considered in a 2D approximation.  They found that the condition for
nonlinearity or the formation of significant trailing shock waves in
the response is that $r_t \left( \Omega/c_s \right) > 0.5,$ where $r_t
= r \left( M_{pl}/M_{*} \right)^{1/3}$ is a multiple of the Roche lobe
radius.  This condition effectively compares the strength of the
protoplanet's gravity to local pressure forces.  For the disk models
considered here, this condition is met for $M_{pl}=10$~M$_{\oplus}$
for all disk mass accretion rates $\dot{M}$ at the inner radii.  Even
for $M_{pl}=1$~M$_{\oplus}$, it is met at 0.1~AU for the higher
accretion rates.  Thus if we wish to consider core mass migration,
nonlinear effects must be considered.  These are expected to lead to a
feedback reaction from the disk which couples the orbital migration to
the viscous evolution of the disk (Lin~\& Papaloizou 1986).  However,
this timescale can also be quite short, particularly for the models
with higher accretion rates.  For example, when $\alpha= 10^{-3}$ and
$\dot M = 10^{-7}$~M$_{\odot}$~yr$^{-1},$ the viscous inflow timescale
is on the order of $10^{4}$~yr at 5~AU.

The characteristic timescale of $10^4-10^5$~yr obtained for migration
and estimated for core formation suggests that cores of several earth
masses form at about 5~AU and migrate inwards to small radii.  In
doing so, they continue to grow.  As long as significant planetesimal
accretion onto the core is maintained during the migration, the
critical core mass $M_{crit}$ remains significant and may not be
attained before the core reaches small radii.  At small radii,
typically smaller than 0.1~AU, the planetesimal accretion rate
decreases because the high temperatures have prevented the formation
of planetesimals there.  The critical core mass is then also reduced
below the actual core mass, so that runaway gas accretion can begin.
Note though that the gas accretion phase may become longer than the
disk lifetime if the mass of the core is too small (see
\S~\ref{sec:discussion}).  However, the build--up of a core massive
enough through the merger of additional incoming cores may enable
giant planets to form within the disk lifetime provided there is
enough gas to supply the atmosphere.

The accretion rate onto a protoplanet migrating in an approximately
circular orbit through a planetesimal swarm with surface density
$\Sigma_p$ can be estimated as:

\begin{equation}
{dM_{pl}\over dt}=2\Sigma_pv_R af. 
\end{equation}

\noindent Here we use a simple two dimensional model appropriate for a
thin planetesimal disk from which accretion occurs onto a large
protoplanet.  The impact target radius of the protoplanet is $a$ and
$v_R$ is the relative velocity in collisions.  We consider the case of
near circular orbits.  Then the relative velocity associated with a
collision is expected to be equivalent to the Keplerian shear across
the Roche lobe radius.  Thus we adopt $v_R= 2r_L\Omega$ as
characteristic induced relative velocity.  The factor $f$ takes
account of other effects such as gravitational focusing, which tends
to increase the collision rate, and any local reduction in $\Sigma_p$,
which would be expected to occur if planetesimals are depleted locally
and a gap tends to form (Tanaka \& Ida 1997).  This might be expected
for very slow migration rates but the total amount of accretion would
be expected to be large in that case.

We take $a=0.01r_L$ as characteristic effective size of the
protoplanet core, this representing the actual physical size of a core
with density 3.2~g~cm$^{-3}$ at 0.75~AU.  At smaller radii this is
larger while at larger radii it is smaller.

In this context we note that there is some uncertainty in the size of
the target radius to be used because there may be a disk of bound
planetesimals.  Further, numerical tests indicate that the results we
present below are not very sensitive to the magnitude of the target
radius used because of the effects of gravitational focusing.  This is
also supported by the results of Kary, Lissauer \& Grenzweig (1993).
They considered the accretion of small planetesimals migrating inwards
under the influence of gas drag by a protoplanet in fixed circular
orbit (see below).  Their results indicate that, for weak gas drag,
which is appropriate for no resonant trapping, the impact probability
typically differs from that assuming a target radius $a=0.01r_L$ by no
more than a factor of about three as $a$ varies between $0.0001r_L$
and $r_L.$ We comment that if the target radius was 6 times the actual
core size, all accretion rates derived here would be underestimates
for migration occurring with $r < 5$~AU.

Thus for a simple estimate we use:

\begin{equation}
{dM_{pl}\over dt}=0.04f\Sigma_p r_L^2 \Omega .
\label{ACCRT}
\end{equation}

\noindent We can estimate the total fraction of the total planetesimal
mass accreted in a migration time $t_{mig}$ to be:

\begin{eqnarray}
{t_{mig} \over \pi \Sigma_p r^2} {dM_{pl} \over dt} & = & \frac{0.04f}{\pi}
\left( {r_L \over r} \right)^2 \Omega t_{mig} 
\nonumber \\
& \sim & 0.01\left({
M_{pl}\over M_{\ast}}\right)^{2/3}\Omega f  t_{mig} .
\label{migration}
\end{eqnarray} 

\noindent Characteristically, we find this fraction to be about 0.1
for $f=1,$ $M_{pl} \sim 10$~M$_{\oplus}$ and $\Omega t_{mig}= 10^{4}$,
which are the expected characteristic values according to Ward
(1997b).  The expected fraction scales only weakly with protoplanet
mass, being proportional to $M_{pl}^{-1/3}$.  But note too that the
above may underestimate the accretion rate because larger relative
velocities may be induced if there are multiple close scatterings.

It is of interest to compare the accretion rate expected from the
above two dimensional model with predictions based on the standard
accretion formula with gravitational focusing for three dimensions
(Lissauer \& Stewart 1993).  This gives:

\begin{equation}
{dM_{pl}\over dt}={\pi a^2 \Sigma_p v_R\over 2h_p}
\left(1+ {2GM_{pl}\over  a v_R^2}\right) ,
\end{equation}

\noindent where $h_p = v_R/\Omega$ is the semithickness of the
planetesimal distribution. Using the same estimate for $v_R$ as above,
this gives (assuming the dominance of the second gravitational
focusing term in the brackets):

\begin{equation}
{dM_{pl}\over dt}={81\pi a  r_L  \Sigma_p\Omega \over 32 } .
\end{equation}

\noindent For the same parameters as used above this gives the same
prediction as equation~(\ref{ACCRT}) with $f=2.$ Both these
expressions and our simulations give consistent results suggesting
that the migrating protoplanet accretes as if it were in a homogeneous
medium without a gap forming in the planetesimal distribution.
Whether such a gap forms should be reliably determined by the two
dimensional calculations.

Given that the disk is expected to contain at least about
8~M$_{\oplus}$ within 5~AU in the early stages (see
\S~\ref{sec:steady} and Table~\ref{tab1}), these estimates indicate
that an accretion rate $\dot{M}_{core}$ of at least about
$10^{-6}$~M$_{\oplus}$~yr$^{-1}$ is likely to be maintained during
orbital migration in the present disk models.  Efficient gas accretion
is then unlikely to start until small radii are reached, at least in
the early phases of the disk lifetime. Note too that the fractional
accretion rate given by equation~(\ref{migration}) is not expected to
increase indefinitely with $t_{mig}$ because of the tendency to form a
gap (Tanaka \& Ida 1997) which would then be expected to cause a
reduction in $f.$

In order to verify the above conclusions, we have performed
simulations of migrating protoplanets with $M_{pl}=10$~M$_{\oplus}$
and migration times, $t_{mig},$ measured at 1~AU, of between $2 \times
10^3$~yr and $10^4$~yr.  We have also considered the case
$M_{pl}=1$~M$_{\oplus}$ and $t_{mig}$ between $2 \times 10^4$~yr and
$10^5$~yr.  The protoplanet was taken to be in a quasi--circular orbit
with $\ln(r)$ decreasing on the specified timescale.  The protoplanet
was assumed to start at 1~AU but the results can be scaled to any
other initial radius in the usual way.  The 256 planetesimals were
initially regularly spaced between 0.6~AU and 0.8~AU, as indicated in
Figure~\ref{fig6}.  The protoplanet was allowed to migrate through
them.  To estimate the fraction of planetesimals accreted, just as
above, we assumed that any particle approaching the protoplanet within
0.01~$r_L$ was accreted.

We remark that the migration was imposed here, being possibly due to
interactions with the disk.  However, a migration mechanism based on
the scattering of planetesimals alone has been noted by Murray et
al. (1998).  Their mechanism requires the protoplanet orbit to have a
non zero eccentricity, but this could be damped by the disk
interaction (Artymowicz~1994).  

For the protoplanet masses and migration rates we consider, we found
significant accretion of planetesimals.  We here present two examples.
The final distribution of the planetesimals after a protoplanet of
1~M$_{\oplus}$ has migrated through them with $t_{mig} = 2 \times
10^4$~yr is indicated in Figure~\ref{fig7}. In this case, about 24\%
of the planetesimals initially present were accreted.  Even more
planetesimals were accreted with slower migration rates.

The final distribution of the planetesimals after a protoplanet of
10~M$_{\oplus}$ has migrated through them with $t_{mig} = 2 \times
10^3$~yr is given in Figure~\ref{fig8}.  In this case, about 23\% of
the planetesimals initially present were accreted.  Given that the
disk models typically contain at least about 8~M$_{\oplus}$ interior
to 5~AU, we conclude that enough accretion occurs during the migration
to prevent gas accretion as long as planetesimals are present.

It is of interest to compare the results of our simulations with those
of Kary et al. (1993). These authors considered the accretion of small
bodies migrating inwards under the influence of gas drag onto a
protoplanet in fixed circular orbit. Although this is not an identical
situation to the one we consider here, it is similar enough to make a
comparison interesting.

An important aspect of these simulations is that gas drag causes
eccentricity damping as well as inward migration.  The eccentricity
damping allows particles to be trapped in resonances such that they
stop migrating, maintaining near circular orbits, with the consequence
that close approaches to the protoplanet are avoided.  The resonant
interaction causes eccentricity growth at a rate governed by the
migration rate. This is because energy and angular momentum transfer
from the protoplanet is in the wrong ratio for keeping the other body
in circular orbit.  Without eccentricity damping, near circular orbits
for the particles could not be maintained (see, for example, Lin \&
Papaloizou 1979b for a discussion).

To relate the migration and damping rates, we consider a particle with
semi--major axis $a_p$ and eccentricity $e$ undergoing resonant
interaction with a protoplanet in a near circular orbit with
semi--major axis $a_{pl}.$ Under these conditions, the Jacobi integral:

\begin{equation}
J_{c}= -{GM_{\ast}\over a_p} \left[ {1\over 2} +\left(
{a_p\over a_{pl}} \right)^{3/2} \sqrt{1-e^2} \right],
\end{equation} 

\noindent is conserved for the particle. Thus, changes to $a_{p}$ and
$e$ are related by:
  
\begin{equation}
{d e^2\over dt} = -\left[ \left( {a_{pl}\over a_{p}}\right)^{3/2} -
\sqrt{1-e^2} \right] \sqrt{1-e^2} {1\over a_p}{d a_p\over dt}.
\label{Jacobi}
\end{equation}

\noindent When an inwardly migrating protoplanet pushes an interior
particle in front of it maintaining a fixed period ratio, just as for
the protoplanet, $ -2 a_p (d a_p/dt)^{-1} = t_{mig}.$ Then
equation~(\ref{Jacobi}) implies that the eccentricity increases.  If
the orbit is to maintain a finite eccentricity, this rate of increase
has to be balanced by the the eccentricity damping or circularization
rate due to dissipative processes such as gas drag.  If the
circularization time is $t_{circ} = -e(de/dt)^{-1},$ an equilibrium
eccentricity can be maintained, such that for small $e$:

\begin{equation}
e^2 = \left|
\left( {a_{pl}\over a_{p}}\right)^{3/2} - 1\right|
{t_{circ}\over t_{mig}}. \label{ecc}\end{equation}

\noindent Equation~(\ref{ecc}) also applies to the case of a fixed
protoplanet orbit and particles migrating inwards due to gas drag.  In
that case $ a_{pl} < a_p,$ and the resonant interaction gives positive
rates of increase for both $a_p$ and $e.$ These are balanced by the
inward migration rate due to gas drag and the orbital circularization
rate respectively.

The process of resonant trapping is accordingly expected to be similar
in the cases of a free particle with inward migrating protoplanet and
particle migrating inwards towards a protoplanet on fixed circular
orbit. But it is important to note that the physical processes causing
migration and circularization may both be very different.

An interpolation of the results of Kary et al. (1993) indicates that
resonant trapping is important for $ t_{mig} > 6\times 10^{4} $~yr for
a 1~M$_{\oplus}$ protoplanet at 1~AU, and $ t_{mig} > 10^{4} $~yr for
a 10~M$_{\oplus}$ protoplanet at 1~AU.  These migration times are
longer than those proposed by Ward (1997b), or the disk viscous
timescale at 1--5~AU in the early stages of the protoplanetary disk
considered here.  The expectation is that resonant trapping will not
be important then.

When there is no resonant trapping, the fraction of bodies accreted is
similar to that found here, ranging between 10 and 40 percent. We
comment that the high rate of eccentricity damping required for
effective resonant trapping is only likely to be obtained for small
bodies. From Kary et al. (1993), when marginal trapping occurs for a
10~M$_{\oplus}$ protoplanet approached by small bodies with $t_{mig}=
10^4$~yr, $ t_{circ} \sim 30$~yr at 1~AU.  As $t_{circ}$ is
proportional to the radius of the body, values less than 30~yr require
the radius of the body to be smaller than 20~m at 1~AU.  But note that
the relative effectiveness of gas drag tends to increase at smaller
radii. Thus resonant trapping, should it occur, is more likely in the
inner regions of the disk.

Equation~(\ref{ecc}) suggests that $t_{mig}$ and $t_{circ}$ should
scale together for marginal resonant trapping occurring with the same
orbital configuration.  Thus a 10~M$_{\oplus}$ protoplanet at 1~AU
migrating inwards with $t_{mig}= 2\times10^5$~yr should resonantly
trap bodies, causing them also to migrate inwards maintaining a fixed
period ratio, if $ t_{circ} < 6 \times 10^2$~yr. We have verified that
trapping occurs when $ t_{circ} = 10^2$~yr.

However, it appears that resonant trapping as a result of gas drag is
unlikely for planetesimals with radii larger than about 10~km and the
accretion rates should then be similar to those found here.
Departures are to be expected only for small bodies at the slowest
migration rates.

\section{Discussion and Summary}

\subsection{Formation of Short--Period Giant Planets}
\label{sec:discussion}

The above suggests that a protoplanetary core formed at about 5~AU
which migrates inwards will not attain the critical core mass, above
which runaway gas accretion starts, before it reaches small radii
$\sim 0.05$--0.1~AU where planetesimals no longer exist.  Runaway gas
accretion onto a small core can then occur at these radii.  However,
if the core is too small, the gas accretion phase may be longer than
the disk lifetime.

\noindent Even for core masses in the range 15--20~M$_{\oplus}$, the
build--up of a massive atmosphere may take a time $\sim 10^6$~yr
(Bodenheimer et al. 1998).  The reason for this is that once the core
starts to accrete a significant atmosphere, energy production occurs
through its gravitational contraction.  The luminosity produced then
slows down the evolution.  An estimate of the evolutionary timescale
at this stage can be obtained by noting that the luminosity should be
equivalent to that required to make the core critical (and hence to
enable runaway gas accretion to begin), assuming it to be produced by
planetesimal accretion. We may thus estimate the time scale as the
Kelvin--Helmholtz time for our already calculated critical core mass
models. This is given by:

\begin{equation}
t_{KH}= |E|/L_{core},
\end{equation}

\noindent where $E$ is the total internal and gravitational energy of
the gas.

\noindent We have calculated $t_{KH}$ in this way for critical cores
at different radial locations in a disk model with $\alpha = 10^{-2}$
and ${\dot M} = 10^{-7}$~M$_{\odot}$~yr$^{-1}$.  The results are
presented in Table~\ref{tab2}.  Typically, we find that $t_{KH} \sim
10^6-10^7$~yr for core masses in the range 10--20~M$_{\oplus}$ for
radii larger than 0.075~AU.  The core masses required to get such a
characteristic timescale increase rapidly interior to 0.06~AU.
However, we note that they decrease as the mass transfer rate through
the disk does.  The fact that fairly large core masses are required to
give evolutionary timescales comparable or less than the expected disk
lifetime means that mergers of additional incoming cores may be
required in order to produce a core of sufficient mass that real
runaway gas accretion may begin.

\placetable{tab2}

Table~\ref{tab1} indicates that, in a disk with $\alpha=10^{-3}$ and
${\dot M} = 10^{-6}$ or $10^{-7}$~M$_{\odot}$~yr$^{-1}$,
40~M$_{\oplus}$ of planetesimals are contained within 1 or 5~AU.
Therefore, the timescale for building--up a core with a mass between
20 and 40~M$_{\oplus}$ at small radii is typically the timescale it
takes for planetesimals to migrate from 1 or 5~AU down to these small
radii.  According to Ward (1997b), the migration timescale of cores of
a few tenths of an earth mass located at 1 or 5~AU is at most
$10^6$~yr in such a disk, and it decreases with increasing core mass.
Therefore, if planetesimals can be assembled into cores of at least a
few tenths of an earth mass at these radii on a reasonably short
timescale, a massive core could be obtained at small radii on a
timescale much shorter than for {\it in situ \/} formation.

\noindent If the disk has $\alpha=10^{-2}$, 40~M$_{\oplus}$ of
planetesimals are contained within 5 or 11~AU.  In this case again,
the migration timescale of cores of a few tenths of an earth mass
located at 5 or 11~AU is about $10^6$~yr, so that the above discussion
still holds.

A massive core can be built--up through the merger of additional
incoming cores either after having stopped at small radii or on its
way down to small radii (where it would still be expected to be
stopped).  The former process resembles that discussed by Ward
(1997a).  The latter scenario would occur if more massive cores, which
migrate faster, overtake less massive cores on their way down.

Supposing that a protoplanetary core massive enough can be built--up on
its way down to small radii and that it continues to rapidly move
inward until it gets interior to the disk inner boundary, it can only
accrete the gas which is in its vicinity, i.e. typically the amount of
gas contained within $\sim 0.05$--0.1~AU.  Since the core is expected
to reach these radii early in the life of the protoplanetary disk,
there may still be an adequate amount of gas there (see
\S~\ref{sec:steady}) for it to build--up a large envelope and become a
giant planet.  However the conditions for that to happen are rather
marginal.

If the protoplanetary core is stopped at some small radius before the
disk is terminated, it might be able to retain contact with disk gas.
In that case it might be able to accrete enough gas supplied from the
outer disk by viscous evolution to build--up a massive atmosphere.

The question arises as to the nature of any process that can halt the
migration.  One might suppose that because the viscous evolution
timescale of the disk increases as the disk gets older, protoplanet
migration, if type~II, gets slower and slower, and may halt altogether
when the disk dissipates (e.g., Trilling et al. 1998).  However, given
that the migration timescale is shorter at smaller radii, if no
mechanism halts it there, very fine 'tuning' would be required to
produce the large fraction of extrasolar planets found on very close
orbits in this way.

The disk may be terminated by a magnetospheric cavity (Lin et. al
1996) such that tidal torques producing inward migration vanish once
the protoplanet enters it.  However, as a result of such an entry,
contact with the disk is lost and further accretion of gas may be
difficult.  Also magnetospheric cavities may not extend up to a few
tenths of an AU, where some of the extrasolar planets have been found
to orbit.

The way in which the migration of a protoplanet would be halted in the
inner regions is not yet clear (e.g., Bodenheimer et al. 1998).  In
this context, we remark that migration rates are usually considered in
relation to a standard $\alpha$ disk model in which the disk midplane
temperature increases inwards. In this situation inward migration is
expected in general (Ward 1986). However, if the inner disk is
terminated through interaction with a stellar magnetic field, physical
conditions may start to differ in the interaction zone where magnetic
field lines penetrate the disk.  Additional energy and angular
momentum transport mechanisms due to a wind for example may start to
become important (see discussion in \S~\ref{sec:steady}).  As a
result, an inward midplane temperature decrease might be produced.  It
may then be possible that migration could be halted such that the
protoplanet retains contact with disk gas.  

In the context of several cores interacting together, we note that
stellar tides are unlikely to provide enough eccentricity damping for
resonant trapping to occur for the migration rates we consider.  The
circularization time resulting from stellar tides is given by
(Goldreich and Soter 1966):
$$
t_{\rm circ} \left( {\rm yr} \right) \sim 2.8 \times
10^{-5} Q {M_{pl} \over M_{\ast}} {P_{\rm o}
\over 1 \; {\rm day}} \left( {a_{pl} \over R_p} \right)^5 .
$$          
Here $P_{\rm o}$ denotes the orbital period, $R_{p}$ the radius of the
planet and $Q$ is the usual tidal $Q$--value.  For a $10$ earth masses
planet orbiting a solar mass at $0.05$~AU, one typically finds $t_{\rm
circ} \sim 3\times 10^4 Q$~yr. As $Q>1,$ this must exceed the
migration times considered here, and so resonant trapping is
unlikely. This is because normally $t_{\rm circ} \ll t_{mig}$ is
required (Kary et al. 1993, Lin \& Papaloizou 1979b and see
eq.~[\ref{ecc}]).  Similar conclusions can be obtained if stellar
tides acting on a Jovian planet are considered (Lin et al. 1998).
 
However, conditions may be different later in the life of the disk
when the viscous evolution time is longer and type~II migration rates
are slower.  A model of the protostellar disk in which giant planets
form at about 5~AU at a later stage in the life of the disk after
about $10^6-10^7$~yr, when the viscous time is of comparable
magnitude, has been considered by Trilling et al. (1998).  In this
situation the planet is able to open a gap and undergo inward type~II
migration on the viscous timescale.  These authors consider outward
torques due to tidal interaction with the central star and Roche lobe
overflow as well as torques due to disk interaction. Assuming disk
dispersal on a similar timescale they are able to produce giant
planets at a range of orbital radii.  These may then undergo orbital
instability, leaving one inner planet and one or several partners at
larger radii (e.g., Weidenschilling \& Marzari 1996; Rasio \& Ford
1996).  In this regard, it is of interest to note that the only system
detected so far which may be multiple, 55~Cnc (Butler et al. 1997),
has a planet at 0.11~AU and maybe another planet beyond 4~AU.

The above scenario thus might be able to produce short period planets
in the late stages of the life of the disk.  In contrast, the
processes which are the focus of this paper result in the short period
planets originating early in the life of the disk.  They would more
likely result in a single planet at $\sim 0.05-0.1$~AU than at
intermediate radii and do not necessarily produce other giant planets
at larger radii as a result of gravitational scattering processes.  In
this respect, the outcome would be in good agreement with the
observations to date.

\subsection{Summary}
\label{sec:summary}

In this paper we have investigated how the critical core mass
associated with a solid protoplanet varies with location in a
protoplanetary disk.  In the past, work has concentrated on {\it in
situ \/} formation, mainly at orbital distances corresponding to the
neighborhood of Jupiter.  However, the importance of orbital migration
has recently been indicated by both observations (see Marcy~\& Butler
1998 and references therein) and theory (see Lin~\& Papaloizou 1993
and references therein; Ward 1997b).  This suggests that the behavior
of the critical core mass as a function of location in and mass
transfer rate through the protoplanetary disk should be considered.

We constructed steady state protostellar disk models with constant
values of $\alpha =10^{-2}$ and $\alpha =10^{-3}$.  A range of
accretion rates varying between $\dot M=10^{-6}$ and
$10^{-9}$~M$_{\odot}$~yr$^{-1}$ were considered.  We calculated
analytic piece--wise power law fits to the curves $\langle \nu \rangle
\left( \Sigma \right)$ obtained from numerical calculations.  These
fits can be used to solve the diffusion equation governing the disk
evolution for a wide range of disk parameters.

We constructed protoplanet models with a solid core and a gaseous
envelope in hydrostatic and thermal equilibrium.  We calculated the
critical core mass, $M_{crit}$, above which the atmosphere cannot
remain in complete equilibrium but must begin to undergo a very rapid
contraction. Where they can be compared, these critical core masses
agree with the crossover masses obtained from more sophisticated time
dependent calculations (Bodenheimer \& Pollack 1986, Pollack et
al. 1996).

We found that, for a fixed core accretion rate, $M_{crit}$ typically
increases by factors of 2--3 between 5 and 0.05~AU for disk parameters
believed to be typical of the early stages of the disk evolution
(i.e. $\alpha=10^{-2}$ and a gas accretion rate $\dot{M} =
10^{-7}$~M$_{\sun}$~yr$^{-1}$).  For a core accretion rate of
$\dot{M}_{core} = 10^{-6}$~M$_{\oplus}$~yr$^{-1}$, the critical core
mass is found to be about 16~M$_{\oplus}$ at 5~AU, in agreement with
Bodenheimer~\& Pollack (1986), and it decreases with $\dot{M}_{core}$.
At radii smaller than about 0.05-0.1~AU, local conditions may not
enable the planetesimals, required to produce the accretion luminosity
to support the atmosphere, to exist. Therefore the critical core mass
is reduced at these radii.
 
We considered these results in the context of the migration of
protoplanetary cores with mass in the range 1--10~M$_{\oplus}$.
According to recent estimates of the effects of tidal interactions
with the disk (Ward 1997b), the migration timescale for such cores may
be between $10^4$ and $10^5$~yr at $5$~AU, being comparable to their
proposed formation time.  In order to investigate whether the
planetesimal accretion would continue under migration, we performed
simulations which indicate that a protoplanet migrating at the
proposed rate is likely to accrete 24\% or more of the planetesimals
initially interior to its orbit.  Thus accretion is likely to maintain
the core luminosity such that attainment of the critical mass does not
occur until small radii $\sim 0.1$~AU are reached, where planetesimals
no longer exist.

Although runaway gas accretion can then begin onto small mass cores at
these small radii, the timescale for building--up a massive envelope
becomes longer than the disk lifetime if the core is too small.
However, cores massive enough can be built--up through mergers of
additional incoming cores on a timescale shorter than for {\it in situ
\/} formation.

The above considerations can lead to the preferential formation of
short--period planets with semi--major axis less than about 0.1~AU, on
a timescale shorter than that derived for {\it in situ \/} planet
formation.

\begin{acknowledgements} 

We acknowledge the Isaac Newton Institute for hospitality and support
during its programme on the Dynamics of Astrophysical Discs, when this
work began. We thank D. Saumon for making his equation of state tables
available to us and P. Bodenheimer for useful discussions.  We also
thank the referee, Jack Lissauer, whose comments helped to improve the
quality of this paper.  CT is supported by the Center for Star
Formation Studies at NASA/Ames Research Center and the University of
California at Berkeley and Santa Cruz, and in part by NSF grant
AST--9618548.

\end{acknowledgements} 


\begin{appendix}


\section{A. Boundary conditions for the disk vertical structure}
\label{appendixA}

We derive here the surface pressure $P_s$ and temperature $T_s$ used
to compute the disk vertical structure (see \S~\ref{sec:boundary}).

To get $P_s$ we rewrite equation~(\ref{dPdz}) under the form:

\begin{equation}
\frac{dP}{d \tau} = - \frac{ \Omega^2 z }{ \kappa} ,
\end{equation}

\noindent where $\tau(z) = \int_0^z \rho \kappa dz $ is the optical
depth.  We then integrate this equation over $\tau$ between the surface
of the disk and infinity, where the pressure is zero.  This leads
to:

\begin{equation}
P_s = \int_{\tau (H)}^{\tau_(\infty)} \frac{ \Omega^2 z}{ \kappa} d \tau .
\label{Ps1}
\end{equation}

\noindent We define the disk surface such that the atmosphere above
the disk is isothermal.  The mass density above the disk then varies
like exp$\left[ - \Omega^2 \left( z^2 - H^2 \right) / \left( 2 c_s^2
\right) \right]$, where $c_s$ is the (constant) sound speed in the
atmosphere.  The main contribution to the integral in
equation~(\ref{Ps1}) comes from values of $z$ starting from $H$ and
extending over a range of a few $c_s/\Omega$ which is significantly
less than $H.$ Thus the integral can be evaluated with sufficient
accuracy by taking $\Omega^2 z$ and $\kappa$ to be constant and equal
to their values at the disk surface.  This leads to:

\begin{equation}
P_s = \frac{\Omega^2 H \tau_{ab}}{\kappa_s} ,
\end{equation}

\noindent where $\tau_{ab}=\int_H^{\infty} d\tau$ is the optical depth
above the disk.

We now calculate the surface temperature $T_s$.  The radiative flux at
the disk surface can be written under the form:

\begin{equation}
F_s = F_+ - F_- ,
\label{Fs2}
\end{equation}

\noindent where $F_+$ and $F_-$ are the radiative vertical fluxes at
the surface of the disk directed respectively along the positive and
negative $z$. In other words, $F_+$ and $F_-$ come respectively from
inside and above the disk. The flux $F_-$ can be written in term of
the background temperature $T_b$ in which the disk is embedded as $F_-
= \sigma T_b^4$. Using equations~(\ref{Fs}) and~(\ref{Fs2}), we then
get:

\begin{equation}
F_+ = \frac{3}{8 \pi} \dot{M}_{st} \Omega^2 + \sigma T_b^4 .
\label{F+1}
\end{equation}

\noindent An other expression of $F_+$ can be obtained by using
the fact that the energy density at the disk surface is given by:

\begin{equation}
E_s = \frac{2}{c} \left( F_+ + F_- \right) ,
\label{Es2}
\end{equation}

\noindent together with the energy equation:

\begin{equation}
\mbox{\boldmath $\nabla$} \cdot {\bf F} = \rho \kappa c \left( a T^4 -
E \right) ,
\label{E}
\end{equation}

\noindent where ${\bf F}$ is the vector representing the flux of
radiative energy, $E$ is the energy density, $c$ is the speed of light
and $a=4 \sigma/c$ is the radiation constant.  In the thin disk
approximation, the temperature gradient in the vertical direction is
much larger than that in the horizontal direction, so that
$\mbox{\boldmath $\nabla$} \cdot {\bf F} \simeq \partial F / \partial
z$. Using the expression~(\ref{dFdz1}) and the relation between $F_-$
and $T_b$, we then get from equations~(\ref{Es2}) and~(\ref{E})
written at the disk surface:

\begin{equation}
F_+ = 2 \sigma T_s^4 - \sigma T_b^4 - \frac{9 \alpha \left(
c_s^2 \right)_s \Omega}{8 \kappa_s} ,
\label{F+2}
\end{equation}

\noindent where $\nu$ has been expressed in term of $\alpha$.  By
comparing equations~(\ref{F+1}) and~(\ref{F+2}) we obtain the
following equation for $T_s$:

\begin{equation}
2 \sigma \left( T_s^4 - T_b^4 \right) - \frac{9 \alpha \left( c_s^2
\right)_s \Omega}{8 \kappa_s} - \frac{3}{8 \pi} \dot{M}_{st} \Omega^2
= 0 .
\end{equation}

\vspace*{0.5cm}

\section{B. Analytical fits of the vertical structure models}
\label{appendixB}
 
To compute the evolution of a non--steady $\alpha$--disk, it is
necessary to solve the diffusion equation~(\ref{aa4}).  To do this,
$\dot{M}_{st}=3 \pi \langle\nu \rangle \Sigma$ has to be specified as
a function of $\Sigma$ at each radius.  We found it convenient to have
an analytic fit to the curves $\dot{M}_{st} \left( \Sigma \right).$
 
We utilize fits in which these curves are approximated by three
different power laws corresponding respectively to the optically thin,
intermediate and optically thick regimes.  The index of these power
laws is independent of the radius $r$ and the parameter
$\alpha$.  However, the multiplicative constant characterizing each of
them does vary with both $r$ and $\alpha$.  We give this dependence
below.
 
Figure~\ref{fig9} shows a schematic plot of the fits $\dot{M}_{st}
\left( \Sigma \right)$. We have represented $\log_{10} \left (
\dot{M}_{st} \right)$ vs. $\log_{10} \left( \Sigma \right)$ with an
arbitrary scale at two different arbitrary radii $r_1$ and $r_2$ such
that $r_2 > r_1$.  When the radius is increased, the point at the
transition between the optically thin and intermediate regimes on this
logarithmic representation moves up along the straight line with
equation is $y=3.1x+c'$.  We now give details of these power laws,
which give a good fit (see below) for $\dot{M}_{st} \le
10^{-4}$~M$_{\sun}$/year and $10^{-5} \le \alpha \le 10^{-1}$.  In the
following expressions, $\dot{M}_{st}$, $r$ and $\Sigma$ are in cgs
units and the logarithms are to base $10.$

In the optically thin regime, as long as $\Sigma \le \Sigma_1$ with
$\log \left( \Sigma_1 \right) = \left( c_1 - c' \right) / 2.1$,
or equivalently $\dot{M}_{st} \le \dot{M}_{st,1}$ with $\log
\left( \dot{M}_{st,1} \right) =(3.1c_1-c')/2.1$, we have:
 
\begin{equation}
\log \left( \dot{M}_{st} \right) = c_1 + \log \left( \Sigma
\right) .
\end{equation}
 
\noindent In the intermediate regime, for $\Sigma_1 \le \Sigma \le
\Sigma_2$ with $\log \left( \Sigma_2 \right) = \left( c_3 - c_2
\right) / 0.9$, or equivalently $\dot{M}_{st,1} \le \dot{M}_{st} \le
\dot{M}_{st,2}$ with $\log \left( \dot{M}_{st,2} \right)
=(2c_3-1.1 c_2)/0.9$, the fit is:
 
\begin{equation}
\log \left( \dot{M}_{st} \right) = c_2 + 2 \log \left( \Sigma
\right).
\end{equation}

\noindent Finally in the optically thick regime, for $\Sigma \ge
\Sigma_2$, or equivalently $\dot{M}_{st} \ge \dot{M}_{st,2}$, we use:
 
\begin{equation}
\log \left( \dot{M}_{st} \right) = c_3 + 1.1 \log \left( \Sigma
\right).
\end{equation}
 
\noindent The parameters $c_1$and $c_3$ are related to $r$ and
$\alpha$ in the following way:
\begin{equation}
\log(c_1) =  0.9360636+0.1195816 \; \log(\alpha) +
\left[ 0.0233002-0.0061733 \; \log(\alpha) \right] \; \log(r) ,
\end{equation}
\begin{equation}
\log(c_3)  =  0.7782080 + 0.0545617 \; \log(\alpha) +
\left[ 0.0366565 - 0.0019087 \; \log(\alpha) \right] \; \log(r) ,
\end{equation}
\noindent whereas $c'$  depends only on $\alpha$:
\begin{equation}
c' = 16.0897161 + 2.0665 \; \log(\alpha) ,
\end{equation}
\noindent and 
\begin{equation}
c_2 = \frac{1.1 c_1 + c'}{2.1} .
\end{equation}
 
\noindent We note that these fits can be used either to compute
$\Sigma$ from $\dot{M}_{st}$ or $\dot{M}_{st}$ from $\Sigma$.

We calculate the error corresponding to these fits by solving the
vertical structure (i.e. calculating $\Sigma$) at 50 different radii
from 0.01 to 100~AU and for 50 values of $\dot{M}_{st}$ between
$10^{-10}$ and $10^{-4}$~M$_{\sun}$~yr$^{-1}$ (both the values of $r$
and $\dot{M}_{st}$ are equally logarithmically spaced).  We then
recalculate $\Sigma$ using the fits and $\dot{M}_{st}$ as an input
parameter.  We find that the average error is 24, 17, 14, 10 and 10\%,
respectively, for $\alpha=10^{-5}$, $10^{-4}$, $10^{-3}$, $10^{-2}$
and $10^{-1}$.  The maximum error is between 43 and 50\% for these
values of $\alpha$.  If alternatively we recalculate $\dot{M}_{st}$
using the fits and $\Sigma$ as an input parameter, the average error
is 36, 22, 18, 13 and 16\% whereas the maximum error is 107, 55, 48,
42 and 103\% for the same values of $\alpha$ from $10^{-5}$ to
$10^{-1}$.  To solve the radial diffusion equation, we need to
calculate $\dot{M}_{st}$ from $\Sigma$ (see \S~\ref{sec:evolution}).
We see that,  except for $\alpha=10^{-5}$ and $10^{-1}$, the fits give a
good approximation.  For $\alpha=10^{-1}$, we can get the maximum and
average errors down to about 50 and 10\%, respectively, by limiting
the calculations to $\dot{M}_{st} \ge 10^{-8}$~M$_{\sun}$~yr$^{-1}$.
For $\alpha=10^{-5}$, the maximum error gets down to 76\% as
$\dot{M}_{st}$ varies between $10^{-8}$ and $3 \times
10^{-5}$~M$_{\sun}$~yr$^{-1}$, while the average error does not change.

\end{appendix}


\begin{table}
\caption[]{Mass of planetesimals contained within a radius $r$. \\
Listed are the radius $r$ of the disk location in AU, $\alpha$, the
gas accretion rate $\dot{M}$ through the disk in M$_{\sun}$~yr$^{-1}$
and the estimate for the  mass of planetesimals 
contained within the radius $r$ given by
$M_p(r)=10^{-2} \times \pi r^2
\Sigma(r)$ in M$_{\oplus}$. \\ }
\begin{tabular}{cccc} 
\tableline
\tableline
$r$ & $\alpha$ & $\dot{M}$ & $M_p(r) = 10^{-2} \times \pi r^2 \Sigma(r)$
\\
(AU) & & (M$_{\sun}$~yr$^{-1}$) & (M$_{\oplus}$) \\ 
\tableline
1   & $10^{-2}$ & $10^{-6}$ &  3.5 \\ 
... & ...       & $10^{-7}$ &  0.6 \\ 
5   & ...       & $10^{-6}$ &  22.8 \\ 
... & ...       & $10^{-7}$ &  7.6 \\ 
1   & $10^{-3}$ & $10^{-6}$ &  30.1 \\ 
... & ...       & $10^{-7}$ &  4.0 \\
5   & ...       & $10^{-6}$ &  138.5 \\ 
... & ...       & $10^{-7}$ &  41.0 \\
\tableline
\end{tabular}
\label{tab1}
\end{table} 

\begin{table} 
\caption[]{Kelvin--Helmholtz timescale for different critical core
masses. \\ The first column gives the radius $r$ of the disk location
in AU, the second column gives the internally generated protoplanet
luminosity as what would be derived from a core accretion rate
$\dot{M}_{core}$ in M$_{\oplus}$~yr$^{-1},$ and the third and fourth
columns give the core mass $M_{core}$ and total mass $M_{pl}$ in
M$_{\oplus}$, respectively.  The fifth column gives the
Kelvin--Helmholtz time $t_{KH}$ in yr. The calculations were performed
for a disk model with $\alpha= 10^{-2}$ and ${\dot M }=
10^{-7}$~M$_{\odot}$~yr$^{-1}.$ \\ }
\begin{tabular}{ccccc}
\tableline 
\tableline
$r$ &  $\dot{M}_{core}$ & $M_{core}$ &  $M_{pl}$ & $t_{KH}$ 
\\ 
(AU) & (M$_{\oplus}$~yr$^{-1}$) & (M$_{\oplus}$) & (M$_{\oplus}$) &
(yr) \\ 
\tableline
0.05 & 10$^{-6}$ & 42.0  & 56.9 & 4.7$\times$10$^{6}$  \\
...  & 10$^{-7}$ & 28.0  & 32.7 & 3.0$\times$10$^{7}$  \\
0.075& 10$^{-6}$ & 21.5  & 28.2 & 1.9$\times$10$^{6}$  \\
...  & 10$^{-7}$ & 15.0  & 18.9 & 1.4$\times$10$^{7}$  \\ 
0.10 & 10$^{-6}$ & 19.5  & 24.8 & 1.4$\times$10$^{6}$  \\
...  & 10$^{-7}$ & 14.0  & 17.7 & 1.2$\times$10$^{7}$  \\  
0.15 & 10$^{-6}$ & 18.5  & 23.9 & 1.4$\times$10$^{6}$  \\
...  & 10$^{-7}$ & 13.0  & 13.9 & 9.1$\times$10$^{6}$  \\
1.0  & 10$^{-6}$ & 17.0  & 20.9 & 9.0$\times$10$^{5}$  \\
...  & 10$^{-7}$ & 12.0  & 14.5 & 7.8$\times$10$^{6}$  \\ 
5.0  & 10$^{-6}$ & 16.0  & 17.3 & 7.5$\times$10$^{5}$  \\
...  & 10$^{-7}$ & 10.5  & 12.5 & 4.8$\times$10$^{6}$  \\  
\tableline
\end{tabular}
\label{tab2}
\end{table}


\begin{figure}
\plotone{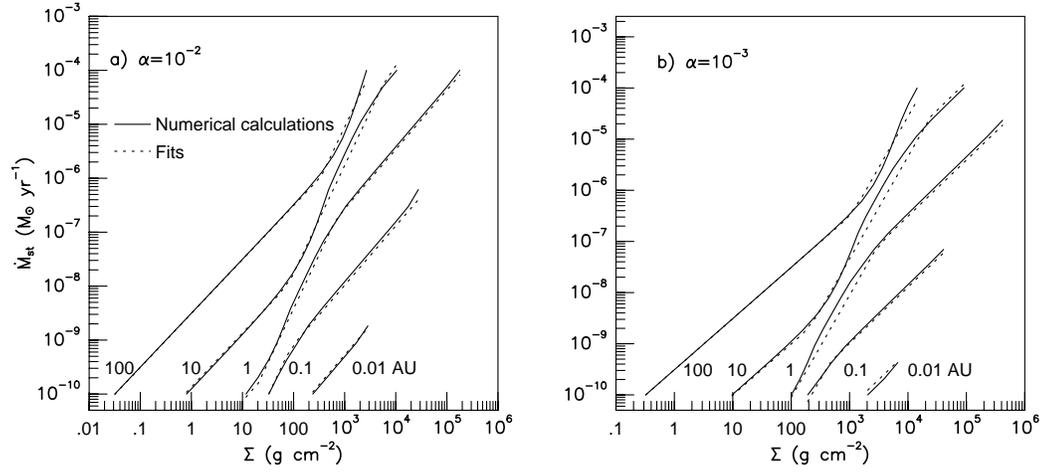}
\caption[]{$\dot{M}_{st}$ in units M$_{\odot}$~yr$^{-1}$ vs. $\Sigma$
in units g~cm$^{-2}$ using a logarithmic scale for $\alpha=10^{-2}$
(a) and $10^{-3}$ (b).  Both the curves corresponding to the numerical
calculations ({\it solid line}) and the fits ({\it dashed line}) are
shown.  The label on the curves represents the radius, which varies
between 0.01 and 100~AU.}
\label{fig1}
\end{figure}       

\begin{figure}
\plotone{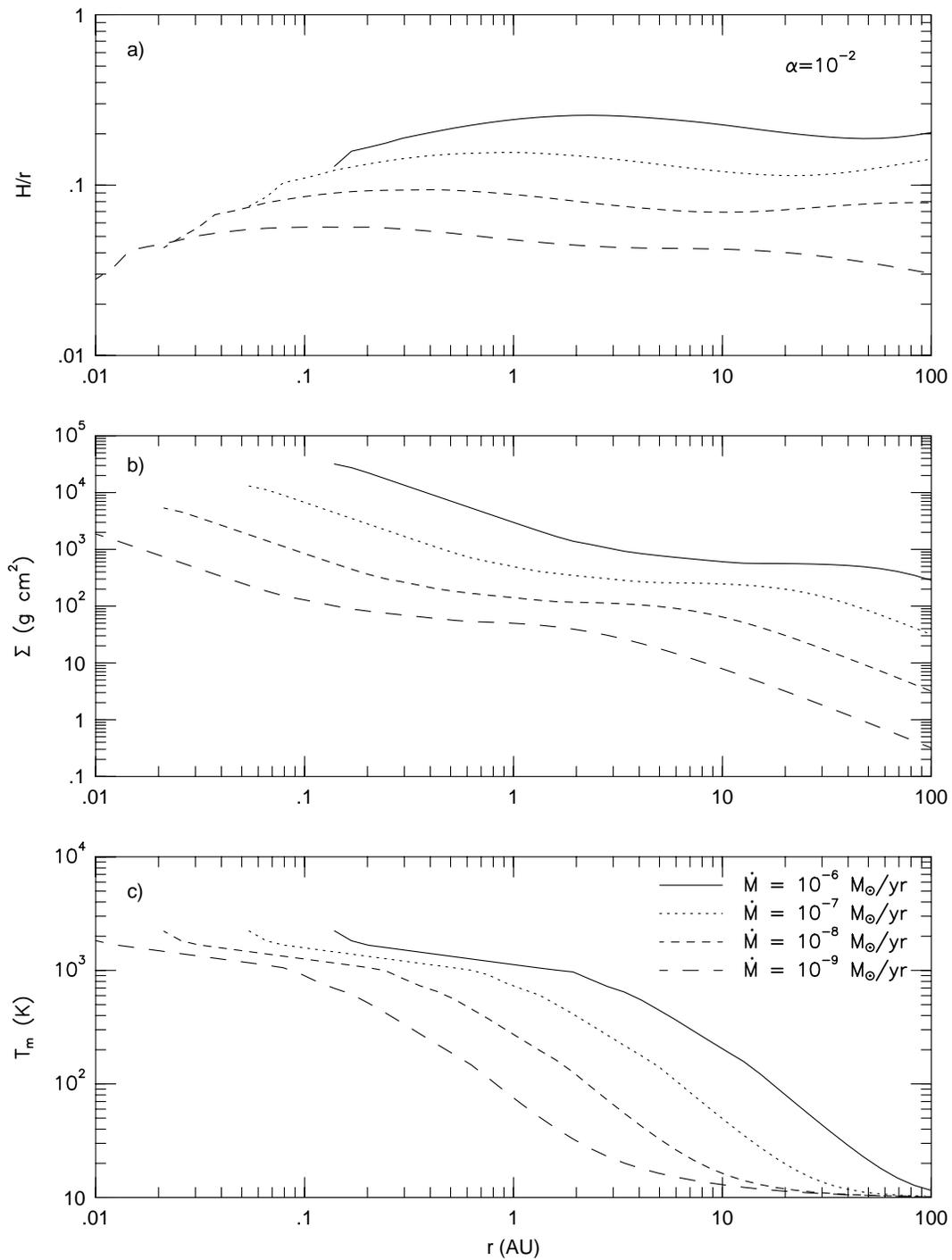}
\caption[]{Shown is $H/r$ (a), $\Sigma$ in units g~cm$^{-2}$ (b) and
$T_m$ in K (c) vs. $r$ in units AU using a logarithmic scale.  In
each plot, the different curves correspond to $\dot{M}_{st} = 10^{-6}$
({\em solid line}), $10^{-7}$ ({\em dotted line}), $10^{-8}$ ({\em
short--dashed line}) and $10^{-9}$ ({\em long--dashed line})
M$_{\sun}$~yr$^{-1}$.  Here $\alpha=10^{-2}$.}
\label{fig1car}
\end{figure}
 
\begin{figure}
\plotone{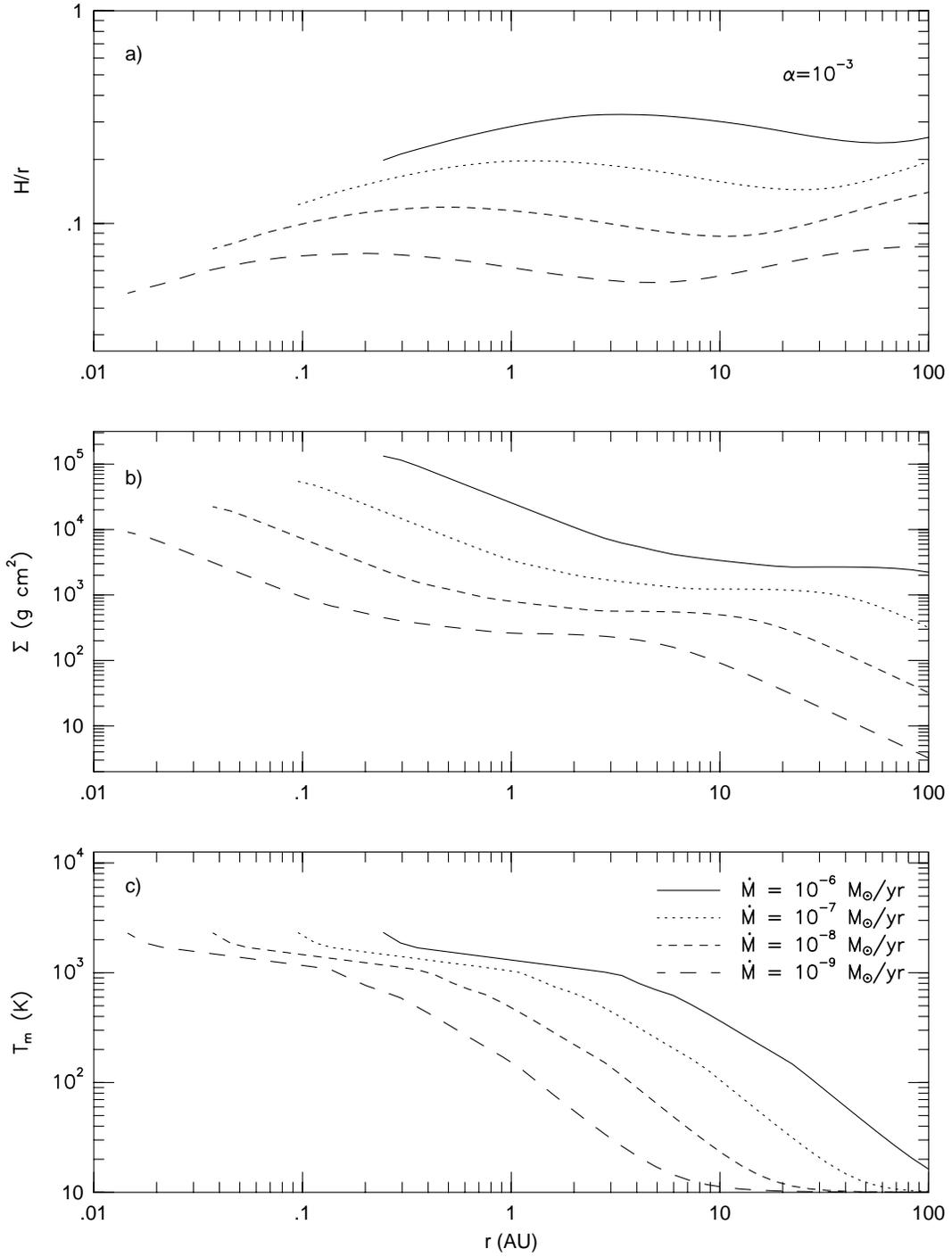}
\caption[]{Same as Figure~\ref{fig1car} but for $\alpha=10^{-3}$}
\label{fig2}
\end{figure}

\newpage
\topmargin -1.5cm

\begin{figure}
\plotone{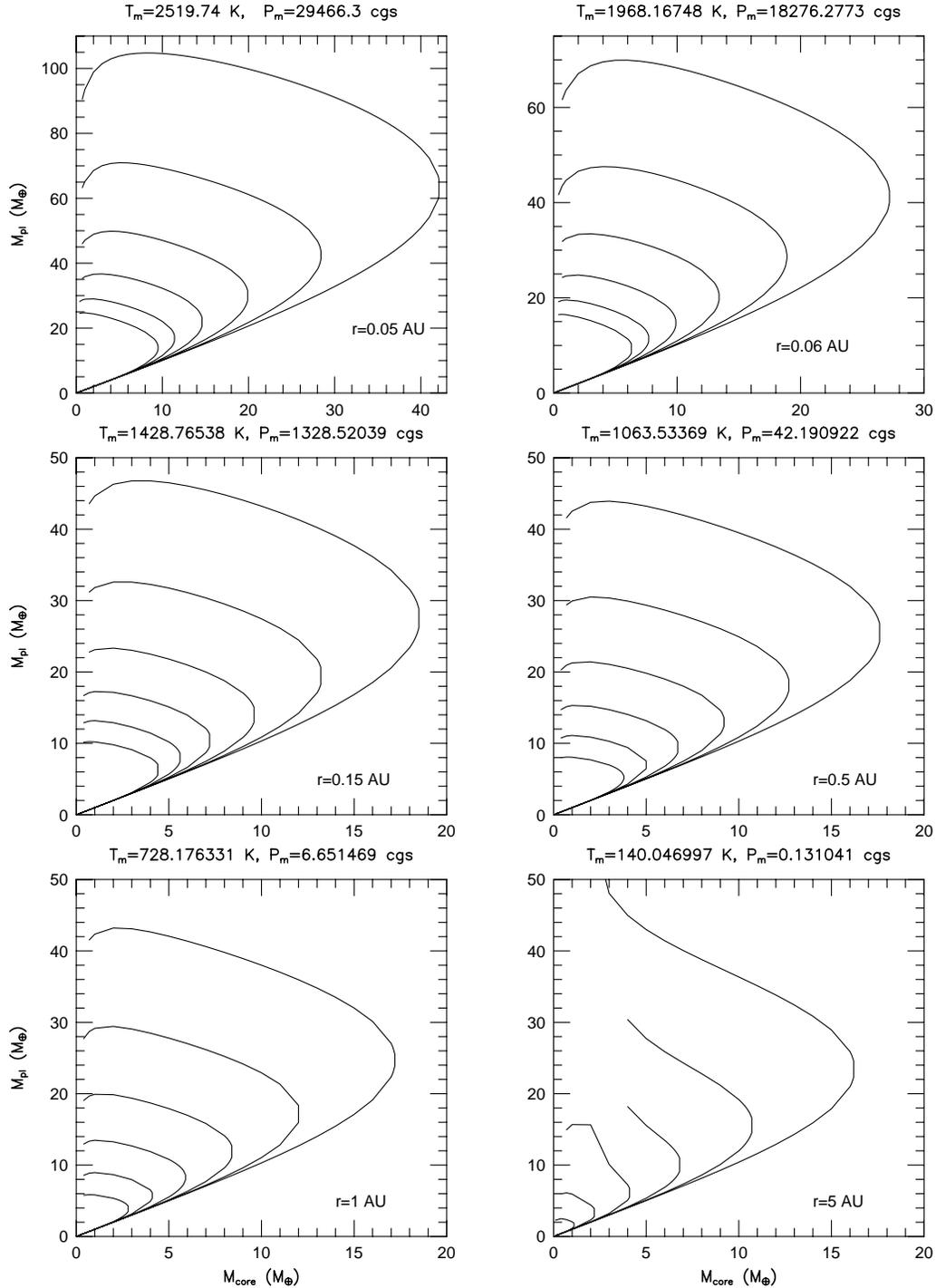}
\caption[]{Plots of total mass, $M_{pl}$ in units M$_{\oplus}$,
vs. core mass, $M_{core}$ in units M$_{\oplus}$, at different
locations $r$ in a steady state disk model with $\alpha=10^{-2}$ and
gas accretion rate ${\dot M} = 10^{-7}$~M$_{\odot}$~yr$^{-1}$.  From
left to right and top to bottom, the frames correspond to $r=0.05$,
0.06, 0.15, 0.5, 1 and 5~AU, respectively.  The midplane temperature
and pressure at these locations are indicated above each frame.  Each
frame contains six curves which, moving from left to right, correspond
to core luminosities derived from planetesimal accretion rates of
$\dot{M}_{core}=10^{-11}$, $10^{-10}$, $10^{-9}$, $10^{-8}$, $10^{-7}$
and $10^{-6}$~M$_{\oplus}$~yr$^{-1}$, respectively.  The critical core
mass is attained when the curves first begin to loop backwards when
moving from left to right.}
\label{fig4}
\end{figure}

\newpage
 
\begin{figure}
\plotone{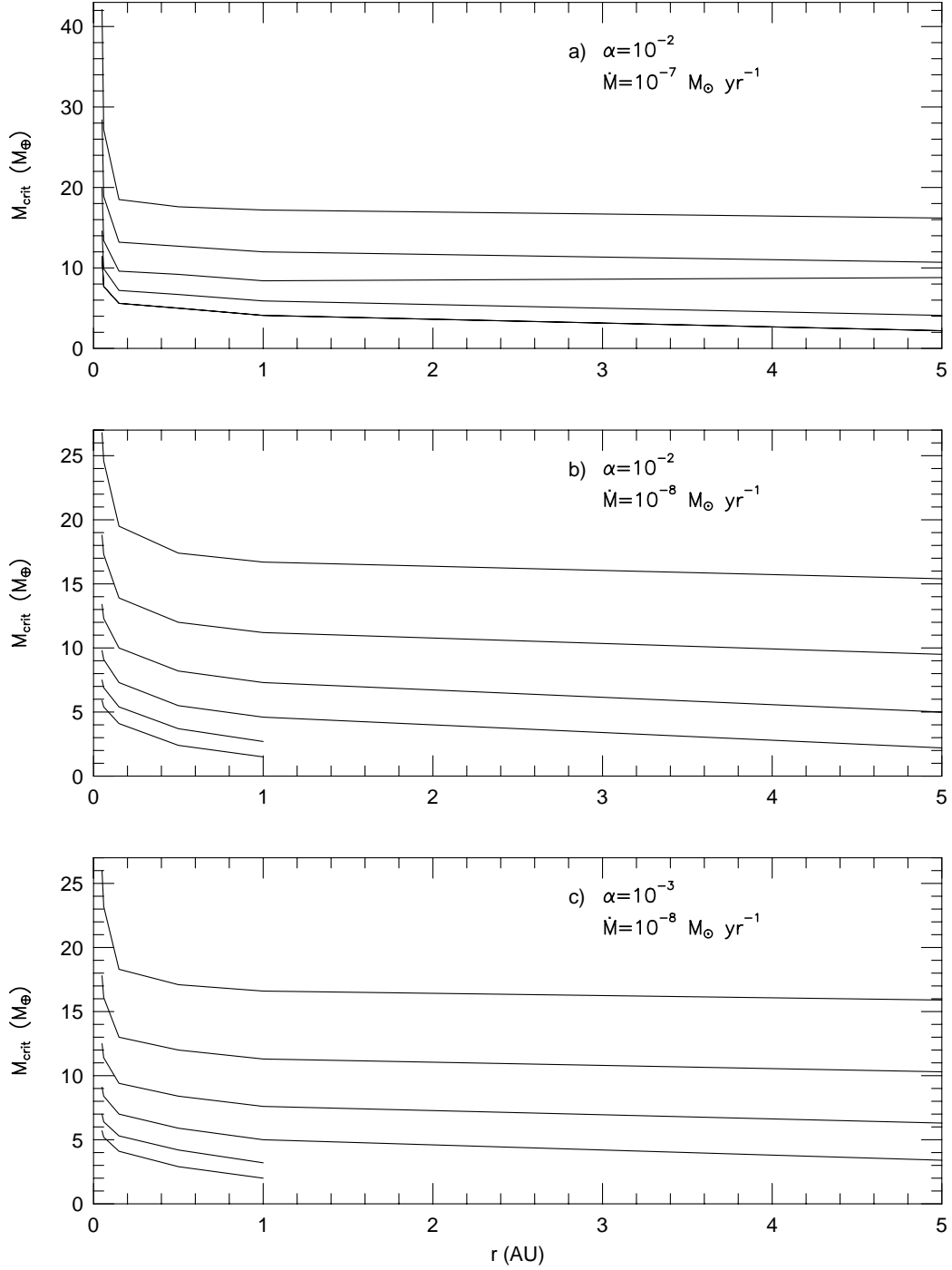}
\caption[]{Critical core mass $M_{crit}$ in units M$_{\oplus}$ vs.
location $r$ in units AU in a steady state disk model with viscous
parameter $\alpha$ and gas accretion rate $\dot{M}$.  The different
plots correspond to $\alpha=10^{-2}$ and $\dot{M} =
10^{-7}$~M$_{\odot}$~yr$^{-1}$ (a), $\alpha=10^{-2}$ and $\dot{M} =
10^{-8}$~M$_{\odot}$~yr$^{-1}$ (b), and $\alpha=10^{-3}$ and $\dot{M} =
10^{-8}$~M$_{\odot}$~yr$^{-1}$ (c).  Each plot contains six curves
which, moving from bottom to top, correspond to core luminosities
derived from planetesimal accretion rates of
$\dot{M}_{core}=10^{-11}$, $10^{-10}$, $10^{-9}$, $10^{-8}$, $10^{-7}$
and $10^{-6}$~M$_{\oplus}$~yr$^{-1}$, respectively.}
\label{fig5}
\end{figure}
 
\begin{figure}
\plotone{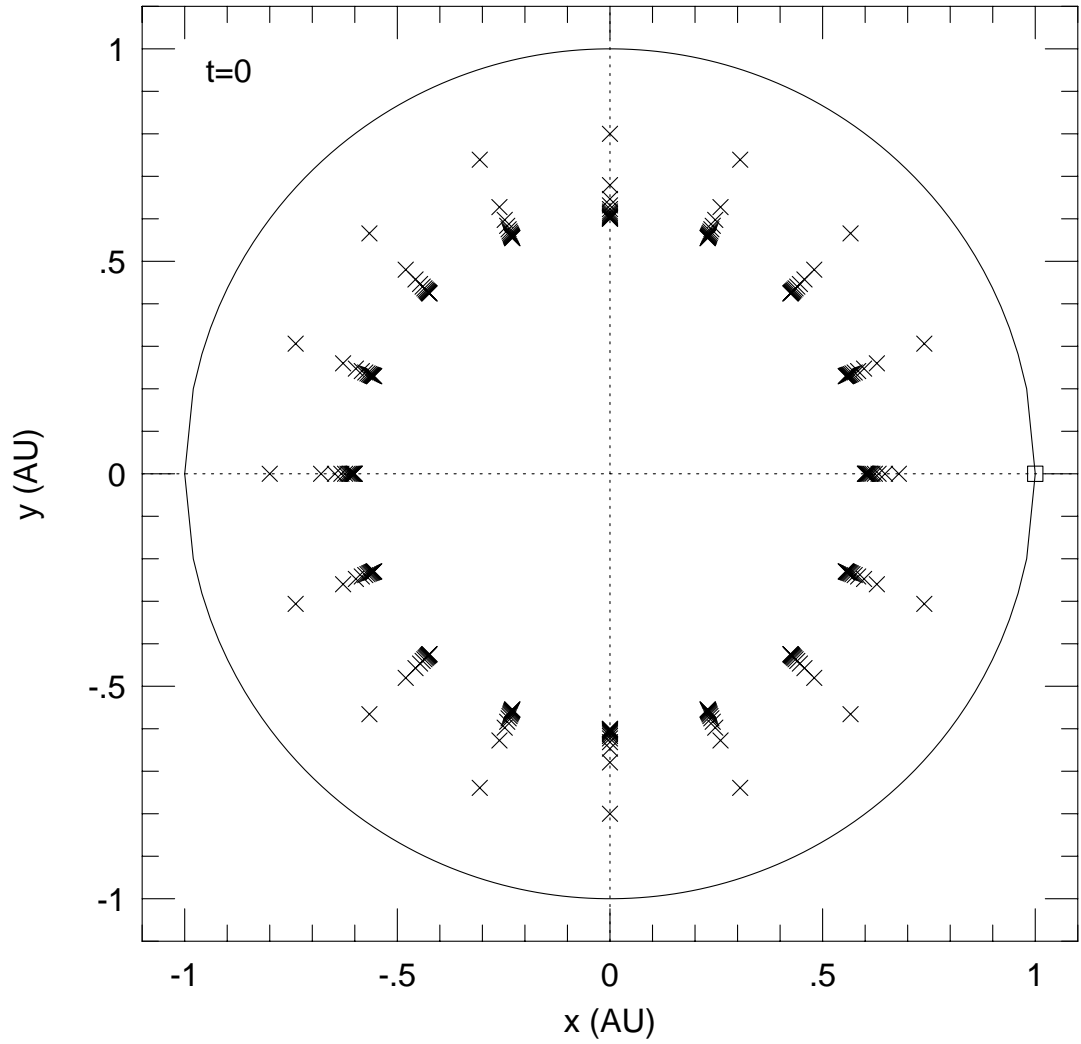}
\caption[]{This shows the initial distribution of planetesimals
(represented by {\it crosses}) between 0.8~AU and 0.6~AU in the
$x$--$y$ plane in units AU.  The initial circular orbit of the
protoplanet at 1~AU is also indicated ({\em solid line}).  The
protoplanet is represented by the {\it open square}.
\label{fig6}}
\end{figure}

\begin{figure}
\plotone{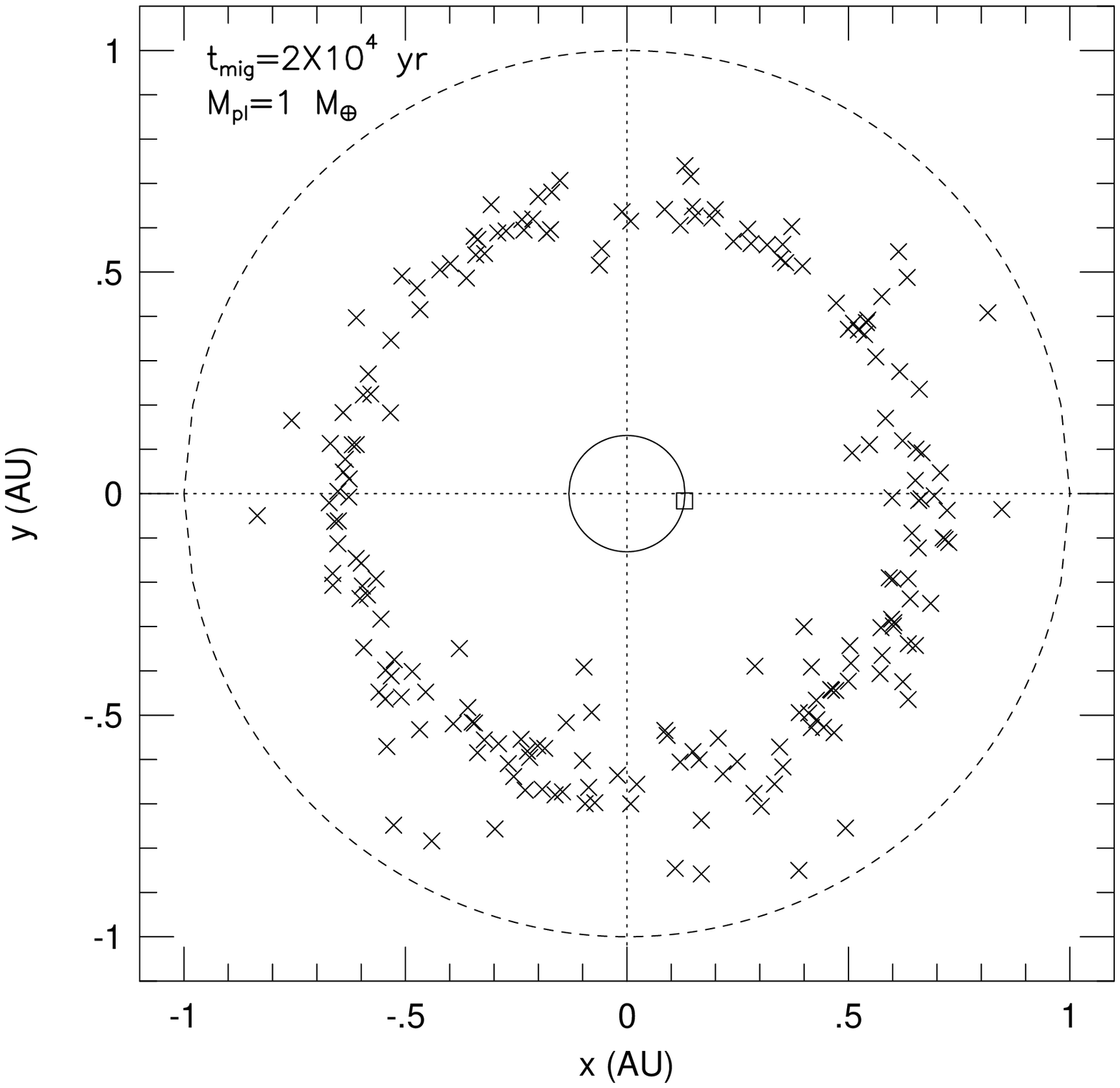}
\caption[]{ The final distribution of planetesimals (represented by
{\it crosses}), in the $x$--$y$ plane in units AU, after a protoplanet
of mass 1~M$_{\oplus}$ has migrated inwards with $t_{mig}= 2 \times
10^4$~yr.  The initial protoplanet orbit is indicated by the outer
circle ({\it dashed line}) and the final one by the inner circle ({\it
solid line}).  The protoplanet is represented by the {\it open
square}.  About 24\% of the planetesimals were accreted.
\label{fig7}}
\end{figure}

\begin{figure}
\plotone{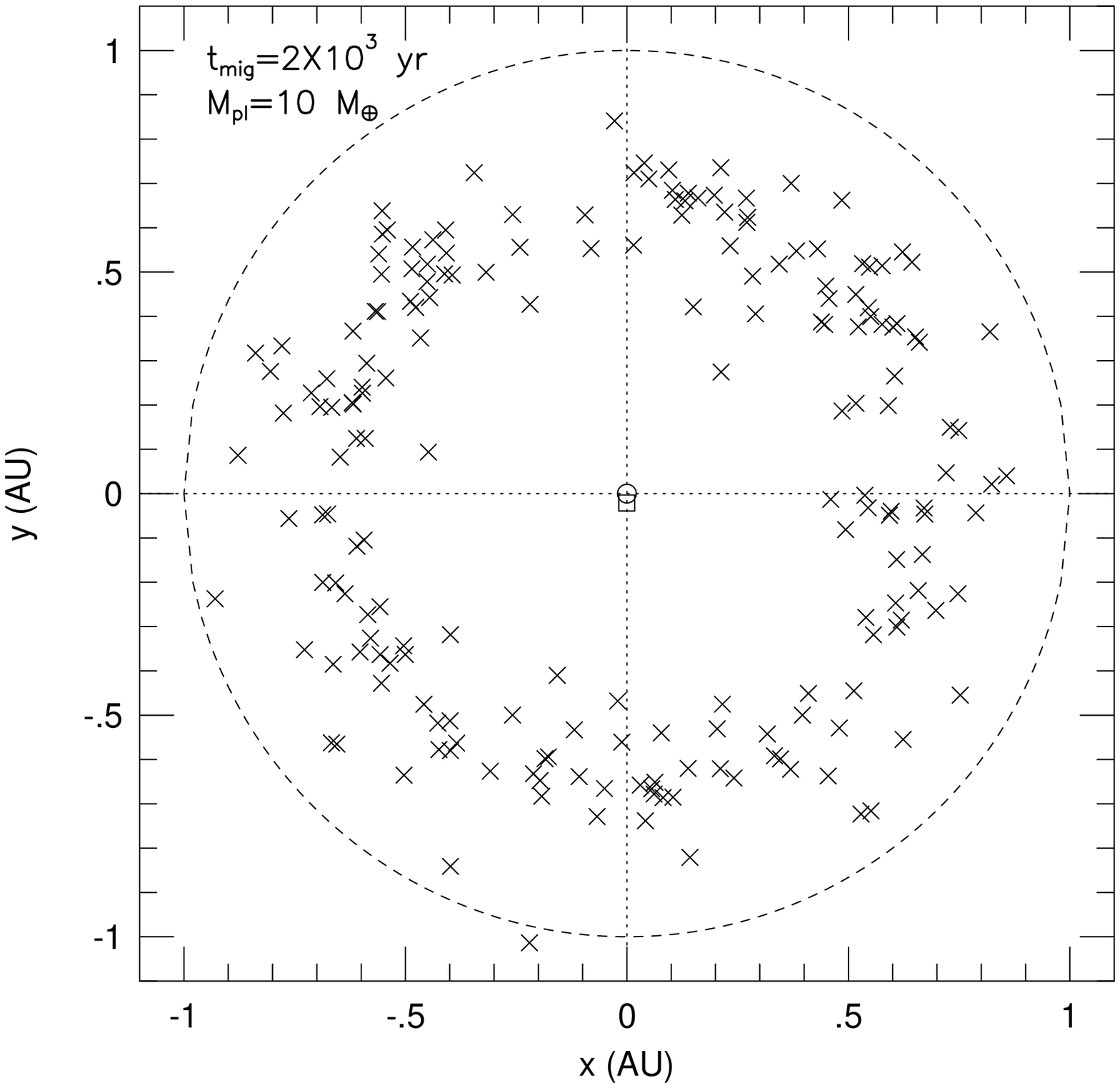}
\caption[]{The final distribution of planetesimals (represented by
{\it crosses}), in the $x$--$y$ plane in units AU, after a protoplanet
of mass 10~M$_{\oplus}$ has migrated inwards with $t_{mig}= 2\times
10^3$~yr.  The initial protoplanet orbit is indicated by the outer
circle ({\it dashed line}) and the final one by the very small inner
circle ({\it solid line}).  The protoplanet is represented by the {\it
open square}.  About 23\% of the planetesimals were accreted.
\label{fig8}}
\end{figure}

\begin{figure}
\plotone{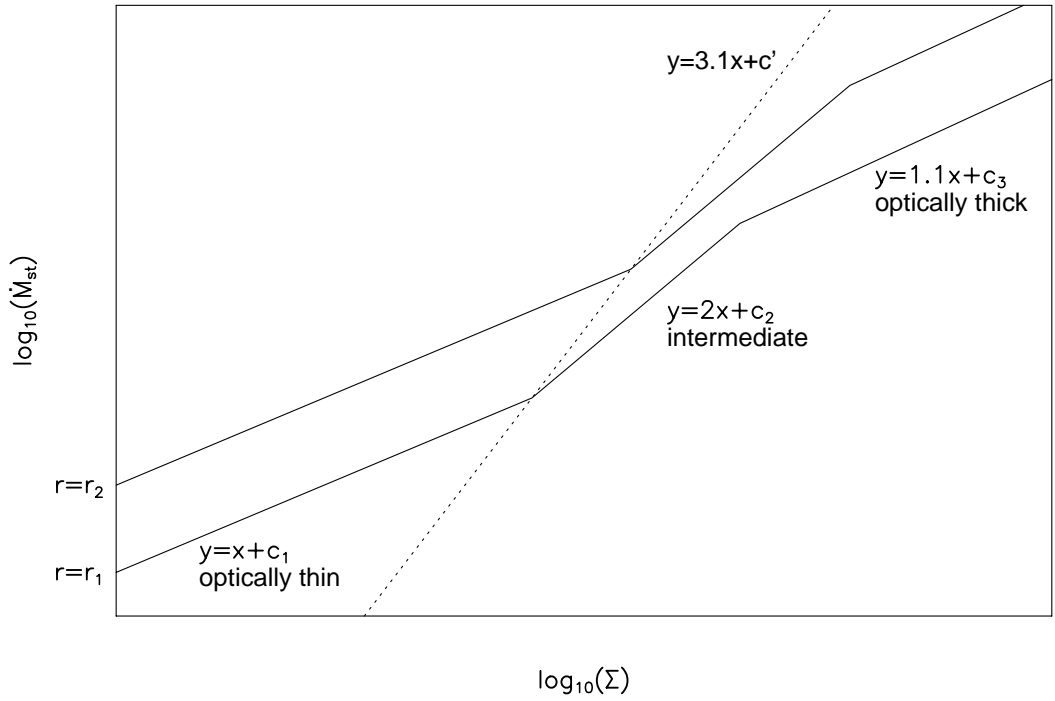}
\caption[]{Schematic plot of the fits $\dot{M}_{st} \left( \Sigma
\right)$. The solid lines represent a fit of the curves $\log_{10}
\left ( \dot{M}_{st} \right)$ vs. $\log_{10} \left( \Sigma \right)$
with an arbitrary scale at two different arbitrary radii $r_1$ and
$r_2$ such that $r_2 > r_1$. The dotted line indicates the separation between the optically thin and intermediate regimes.}
\label{fig9}
\end{figure} 

\end{document}